\documentclass{CLST}
\usepackage{amsmath,amssymb,amsfonts}
\usepackage{graphicx}

\usepackage{textcomp}
\usepackage{xcolor}
\usepackage{threeparttable}
\usepackage{multirow}
\usepackage{booktabs}
\makeatletter 
  \newcommand\figcaption{\def\@captype{figure}\caption} 
  \newcommand\tabcaption{\def\@captype{table}\caption} 
\makeatother
\begin{document}
%\oa
%%%%%%%%%%%%%%%%%%%%%%%%%%%%%%%%%%%%%%%%%%%%%%%%%%%%%%%
%%% 
\ArticleType{AI and Security}
%\SpecialTopic{}
\Year{2021}
\Month{}
\Vol{}
\No{}
\DOI{}
\ArtNo{}
\ReceiveDate{}
\ReviseDate{}
\AcceptDate{}
\OnlineDate{}
%%%%%%%%%%%%%%%%%%%%%%%%%%%%%%%%%%%%%%%%%%%%%%%%%%%%%%%
%
\title{EvilModel 2.0: Bringing Neural Network Models into Malware Attacks} {EvilModel 2.0}
\author[]{Zhi Wang}{}
\author[]{Chaoge Liu}{}
\author[]{Xiang Cui }{}
\author[]{Jie Yin}{}
\author[]{Xutong Wang}{}
\abstract{Security issues have gradually emerged with the continuous development of artificial intelligence (AI). Earlier work verified the possibility of converting neural network models into stegomalware, embedding malware into a model with limited impact on the model's performance. However, existing methods are not applicable in real-world attack scenarios and do not attract enough attention from the security community due to performance degradation and additional workload. Therefore, we propose an improved stegomalware EvilModel. By analyzing the composition of the neural network model, three new methods for embedding malware into the model are proposed: MSB reservation, fast substitution, and half substitution, which can embed malware that accounts for half of the model's volume without affecting the model's performance. We built 550 EvilModels using ten mainstream neural network models and 19 malware samples. The experiment shows that EvilModel achieved an embedding rate of 48.52\%. A quantitative algorithm is proposed to evaluate the existing embedding methods. We also design a trigger and propose a threat scenario for the targeted attack. The practicality and effectiveness of the proposed methods were demonstrated by experiments and analyses of the embedding capacity, performance impact, and detection evasion. }

\maketitle

\section{Introduction}\label{sec:intro}
In recent years, artificial intelligence (AI), represented by deep neural network, had several achievements in various fields and has become a key technology for promoting economic development. However, AI has security issues. Research~\cite{KaloudiL20,OffensiveAI21} has demonstrated how attackers can use AI to enhance all stages of an attack, thereby making the attacks more intelligent, concealed, and efficient. For example, DeepLocker~\cite{Kirat18} introduced a neural network model into targeted attacks and found the target covertly using the model. DeepC2~\cite{wang2020deepc2} uses a neural network in botnet command and control (C2), delivering commands in a block-resistant manner.

With the continuous improvement in the AI industry, AI-powered attacks are more prevalent~\cite{MBrundage18}. Security vendor Beyond Trust~\cite{BeyondTrust20} pointed out that the proliferation of AI weaponization will bring significant challenges to cybersecurity. Gartner~\cite{gartner19} stated that, by 2022, 30\% of cybersecurity issues would be related to AI security. Fortinet~\cite{Fortinet20} also believes that the weaponization of AI will become a trend. Therefore, it is necessary to study AI-powered threats to better understand and defend against them. 

Recent research has shown that neural network models can be used as malware carriers for attacks. Stegomalware called StegoNet~\cite{Liu20stegonet} was proposed to embed malware into neural network models. The parameters in the models were replaced or mapped with malware bytes. Meanwhile, the model's performance was maintained because of the complexity and fault tolerance of the models. 
One potential attack scenario of StegoNet is to launch supply chain pollution through ML markets~\cite{BVLC21zoo}. With the rise of Machine Learning as a Service (MLaaS)~\cite{aws21, gcloud21, azure21} and the open machine learning market, attackers can propagate polluted models through the MLaaS provider and ML market.

The strength of hiding malware in the neural network models are:
i) Detection evasion. After hiding in the models, the malware cannot be disassembled, and its characteristics cannot be extracted. Therefore, it can evade detection.
ii) Concealment. Because of the redundant neurons and excellent generalization ability, the modified models can maintain their original performance without causing anomalies.
iii) High capacity. Normally, neural network models are large and can embed large malware.
iv) Independence. This method does not rely on system vulnerabilities. Malware-embedded models can be delivered through model update channels from the supply chain or alternative ways that do not attract the end-users attention.
v) Universality. As neural networks become more popular, this method will become universal in spreading malware in the future.

Based on the advantages summarized above, we believe that this attack can be used to spread malware. However, as a classic design, StegoNet still has some deficiencies in real-world scenarios from the attacker's perspective, resulting in insufficient attention from the security community. First, StegoNet has a low embedding rate (defined as a malware/model size). In StegoNet, the upper bound of the embedding rate without accuracy degradation is $\sim$15\%, which is insufficient for embedding large malware into medium- or small-sized models. Second, the StegoNet methods have a significant impact on the model's performance. The accuracy of the models decreased significantly with the malware size increased, particularly for small-sized models. It renders StegoNet unsuitable for small models. Additionally, StegoNet requires extra effort to embed or extract malware, such as extra training and index permutation, making it impractical.

To raise attention to this emerging threat, we conduct further studies and propose new malware embedding methods. We embedded the malware into neural network models with a high embedding rate and low performance impact. We first analyzed the composition of neural network models and studied how to embed the malware and how much malware can be embedded. Based on the analysis, we propose three embedding methods, MSB (most significant byte) reservation, fast substitution, and half substitution, to embed malware. We embedded 19 malware samples in ten mainstream neural network models and analyzed their performance to demonstrate the feasibility of the proposed method. We propose an evaluation algorithm that combines embedding rate, performance impact, and embedding effort to evaluate the existing methods. To demonstrate the potential threat of this attack, we present a possible attack scenario and study the embedding capacity of neural network models.

The contributions of this paper are summarized as follows:
\begin{itemize}
\item We propose three methods to embed malware into neural network models with a high embedding rate and low performance losses. We built 550 malware-embedded models using ten mainstream models and 19 malware samples and evaluated their performances on ImageNet.
\item We propose a quantitative method combining the embedding rate, the model performance impact, and the embedding effort to evaluate and compare the existing methods.
\item We design a trigger and present a potential threat of the proposed attack. We trained a model to identify targets covertly and embedded WannaCry into the model. We use the trigger to activate the extraction and execution of the malware.
\item We further study the neural network model's embedding capacity and analyze the relationship between the model structure, network layer, and the performance impact.
\item Some possible countermeasures were also proposed to mitigate this kind of attack.
\end{itemize}

We note that a shorter version of this study appeared in IEEE ISCC 2021~\cite{Wang2021EvilModel}. The previous version focuses on the proof-of-concept, and this version concentrates on experimentation, evaluation, and application. The additions in this manuscript include i) two more embedding methods that have increased the embedding rate from 20.73\% to 48.52\% without affecting the model performance, making this method possible to use in practice; ii) a comprehensive experiment with 550 EvilModels using more malware samples and models; iii) two additional methods and experiments on evasiveness; iv) a quantitative method to evaluate the existing methods; v) a trigger to activate the malware in a real-world attack; vi) more detailed countermeasures with experiment; and vii) more discussion on the experiments and application.

The rest of this study is organized as follows. Section~\ref{sec:background} describes the background and related work in this study. Section~\ref{sec:method} presents the methodology used to embed malware. Section~\ref{sec:imple} describes the experiment and evaluation of the proposed methods. Section~\ref{sec:app} presents a case study of a potential threat. Section~\ref{sec:em} presents an investigation of the embedding capacity. Section \ref{sec:discuss} discusses possible countermeasures and future work. Section \ref{sec:conclusion} concludes the study.

\section{Background and Related Work}\label{sec:background}

\subsection{Stegomalware and Steganography}
Stegomalware is an advanced malware that uses steganography to avoid detection. Malware is concealed in benign carriers such as images, documents, and videos. A typical steganography method is image-based LSB (least significant bit) steganography~\cite{lsb07}. An image contains pixels with values from 0 to 255. When expressed in binary form, LSBs have little effect on the picture's appearance. Therefore, they can be replaced by secret messages. Thus, the messages are hidden in the images. Due to its low channel capacity, this method is unsuitable for embedding large-sized malware.

With the increasing popularity of AI, neural networks have been applied in steganography. Volkhonski et al.~\cite{VolkhonskiyNB19} proposed a GAN-based image steganography method SGAN (steganographic generative adversarial networks). A carrier image is generated from random noise. The generation network optimizes the network parameters through the feedback of the discriminative and  steganalysis network so that the generated image looks natural after embedding information. 
Zhang et al.~\cite{ZhangZCLLY18} proposed an adversarial samples-based image steganography method that resists steganalysis. Adversarial samples can interfere with the judgment of the neural network by making some small changes to the original image that are invisible to the human eye. Using this feature, adding small changes to the carrier image can make it masquerade as a natural image and fool the steganalyzer.
These methods are mainly applied to image steganography.

\subsection{StegoNet}
StegoNet~\cite{Liu20stegonet} proposes to covertly deliver malware to end devices by malware-embedded neural network models from the supply chain, such as the DNN model market, MLaaS platform, etc. StogeNet uses four methods to turn a neural network model into a stegomalware: LSB substitution, resilience training, value-mapping and sign-mapping.

\textbf{LSB substitution.} Neural network models are redundant and fault-tolerant. By taking advantage of the sufficient redundancy in neural network models, StegoNet embeds malware bytes into the models by replacing the least significant bits of the parameters. For large-sized models, this method can embed large-sized malware without the performance degradation. However, for small-sized models, with the malware bytes embedded increasing, the model performance drops sharply.

\textbf{Resilience training.} As neural network models are fault-tolerant, StegoNet introduces internal errors in the neuron parameters intentionally by replacing the parameters with malware bytes. Then StegoNet ``freezes'' the neurons and retrains the model. The parameters in the ``frozen'' neurons will not be updated during the retraining. There should be an extra ``index permutation'' to restore the embedded malware. Compared with LSB substitution, this method can embed more malware into a model. The experiments show that the upper bound of the embedding rate for resilience training without accuracy degradation is $\sim$15\%. There is still a significant impact on the model performance, although retraining is performed to restore the performance.

\textbf{Value-mapping.} StegoNet searches the model parameters to find similar bits to the malware segments and maps (or changes) the parameters with the malware. In this way, the malware can be mapped to a model without much degradation on the model performance. However, it also needs an extra permutation map to restore the malware. Also, in this way, the embedding rate is lower than the methods above.

\textbf{Sign-mapping.} StegoNet also maps the sign of the parameters to the malware bits. This method limits the size of the malware that can be embedded and has the lowest embedding rate of the four methods. Also, the extra permutation map will be huge, making this method impractical.

The common weaknesses of the above mentioned methods are i) they have a low embedding rate, ii) they have a significant impact on the model performance, and iii) they need extra efforts in the embedding works, mainly index permutation or permutation map. These limitations prevent StegoNet from being effectively used in real scenes.

%EvilModel~\cite{wang2021evilmodel} proposes \textbf{fast substitution} to embed the malware in the model by replacing the neurons. EvilModel analysis the component and distribution of the parameters in a neuron, and replaces the whole neurons with totally different malware-embedded numbers. According to the parameter value, each parameter in a neuron is replaced with three malware bytes and one prefix byte 0x3C or 0xBC. It has a higher efficiency to embed the malware but also a greater impact on the performance of the model. EvilModel also explores how much malware can be embedded in a model and which net layer is more robust to embed malware.

\subsection{DeepLocker}
Kirat et al.~\cite{Kirat18, DLPatent20} proposed a covert-targeted attack scenario known as DeepLocker. DeepLocker uses the environmental attributes of the target (e.g., face, voice, geolocation, and motion) to train a neural network model (we take the faces as an example). The model generates a steady cryptographic key as the output when the target face is the input. The key encrypts malicious payloads. DeepLocker delivers malware-bundled video software to a victim's environment through the supply chain. When the software is running, the malware captures the facial images and inputs them into the neural network model. When the target uses the software, the model recognizes it and outputs the crypto key; otherwise, the output is meaningless numbers. Subsequently, the malicious payload can be decrypted and executed. Security analysts cannot obtain the correct key from a model when the target that decrypts the payload is unknown. Therefore, they are unaware of the malicious payload's intent. DeepLocker transforms the traditional way of identifying targets from an ``if-then'' mode to a black-box mode.

In the implementation, the attackers first need to train a face-recognizing model $M_1$. They can alternatively use pre-trained models as $M_1$. When the model is ready, if the inputs are the faces from the same person, $M_1$ will generate similar but not identical feature vectors. The vectors are used to build another model $M_2$. When training $M_2$, the input is the feature vectors of the target from $M_1$, and the output is a pre-defined cryptographic key. The training of $M_2$ requires approximately ten face images from the target. When $M_2$ is ready, the attackers connect $M_1$ and $M_2$ and obtain the final model $M$ (i.e., $M=M_2(M_1)$). When the input is the target, $M$ outputs the crypto key. In this way, the trigger condition and the target information are hidden in the neural network model, and the malware's intent is hidden in the encrypted payload. The deficiency of DeepLocker is that the encrypted payload is an anomalous content in the victim environment. It reduces the concealment of the attack.

\subsection{Malicious use of AI}
Malware is constantly exploring new means of concealment and enhancement, such as online social networks~\cite{Pantic15, FireEye15}, encrypted DNS queries like DNS over HTTPS/TLS~\cite{Patsakis20dns}, Bitcoin blockchain~\cite{Pony19, WangS20}, and InterPlanetary File System (IPFS)~\cite{PatsakisC19, IPStrtom19}.
As AI exceeds many traditional methods in various fields, it is possible to utilize AI to carry out network attacks that are more difficult to defend.
In 2018, 26 researchers~\cite{MBrundage18} from different organizations warned against the malicious use of AI. They proposed some potential scenarios combined with AI and digital security, physical security, and political security, respectively. At the same time, AI-powered attacks are emerging. 

For preparing an attack, Seymour et al.~\cite{SpearTwitter16} proposed a highly targeted automated spear phishing method with AI. High-value targets are selected by clustering. Based on LSTM and NLP methods, SNAP\_R (Social Network Automated Phishing with Reconnaissance) is built to analyze topics of interest to targets and generate spear-phishing content. The contents are pushed to victims by their active time. Hitaj et al.~\cite{PassGAN19} proposed PassGAN to learn the distributions of real passwords from leaked passwords and generate high-quality passwords. Tests from password datasets show that PassGAN performs better than other rule- or ML-based password guessing methods.

For covert communication, Rigaki et al.~\cite{RigakiG18} proposed using GAN to mimic Facebook chat traffic to make C\&C communication undetectable. Wang et al.~\cite{wang2020deepc2} proposed DeepC2 that used neural network to build a block-resistant command and control channel on online social networks. They used feature vectors from the botmaster for addressing. The vectors are extracted from the botmaster's avatars by a neural network model. Due to the poor explainability and complexity of neural network models, the bots can find the botmaster easily, while defenders cannot predict the botmaster's avatars in advance.

For detection evasion, MalGAN~\cite{HuT17} was proposed to generate adversarial malware that could bypass black-box machine learning-based detection models. A generative network is trained to minimize the malicious probabilities of the generated adversarial examples predicted by the black-box malware detector. More detection evasion methods~\cite{EvadePE18, Yuan20, Wang21} were also proposed after MalGAN.

AI-powered attacks are emerging. Due to its powerful abilities on automatic identification and decision, it is well worth the effort from the community to mitigate this kind of attack once they are applied in real life.

\section{Methodology}\label{sec:method}
In this section, we introduce methodologies for embedding malware into a neural network model.

\subsection{Analysis of the neurons}\label{sec:nn}

\subsubsection{Neurons in a Network} 
\begin{figure}
\centering
\includegraphics[width=0.6\linewidth]{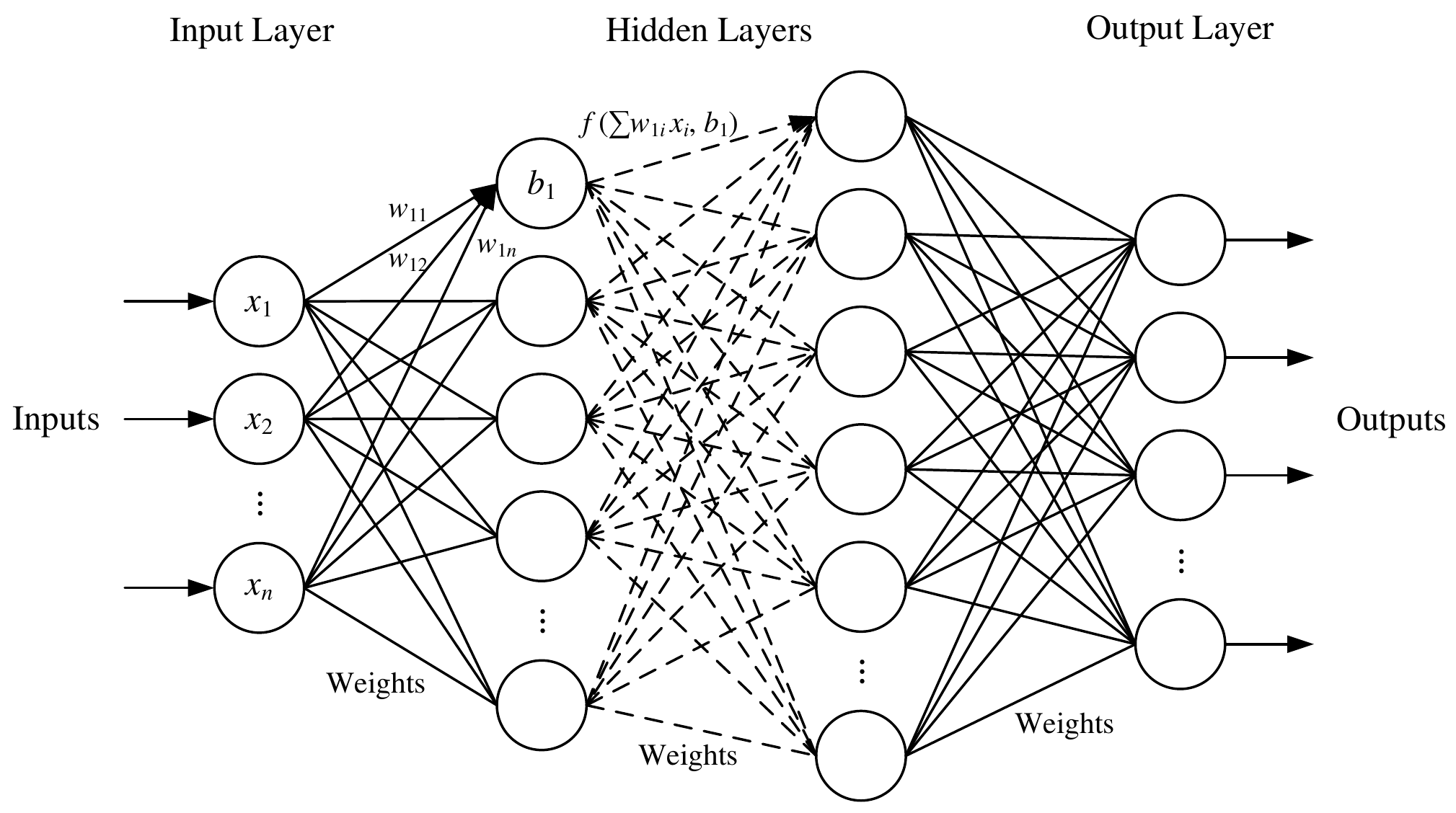}
\caption{Basic structure of neural network models}
\label{fig:mlmodel}
\end{figure}

A neural network model typically has an input layer, one or more hidden layer(s), and an output layer, as shown in Fig.~\ref{fig:mlmodel}. The input layer receives external signals and sends them to the hidden layer of the neural network through the input layer neurons. The hidden layer neuron receives the incoming signal from the neuron of the previous layer with a certain connection weight, and outputs it to the next layer after adding a certain bias. There may be different numbers of neurons in different hidden layers of a neural network. The last layer is the output layer. It receives incoming signals from the hidden layer and processes them to obtain the neural network's output.

Suppose a hidden layer has $m$ neurons and the previous layer has $n$ neurons. A neuron in the hidden layer has connection weight $w_i$ for each input signal $x_i$ from the previous layer. Assume that all inputs of the neuron $\mathbf{x}=(x_1,x_2,...,x_n)$, and all connection weights $\mathbf{w}=(w_1,w_2,...,w_n)$, where $n$ is the number of input signals (i.e. the number of neurons in the previous layer). A neuron receives the input signal $\mathbf{x}$ and calculates $\mathbf{x}$ with the weights $\mathbf{w}$ using matrix operations. Subsequently, a bias $b$ is added to fit the objective function. The output of the neuron is $y=f(\mathbf{wx}, b)=f(\sum_{i=1}^n w_ix_i, b)$. We can see that each neuron contains $n + 1$ parameters, that is, $n$ connection weights (the number of neurons in the previous layer) and one bias. Therefore, a neural layer with $m$ neurons has $m(n+1)$ parameters. In mainstream neural network frameworks (PyTorch, TensorFlow, etc.), each parameter is a 32-bit floating-point number. Therefore, the size of parameters in each neuron is $32(n+1)$ bits, which is $4(n+1)$ bytes, and that of the parameters in each layer is $32m(n+1)$ bits, which is $4m(n+1)$ bytes. Note that the size of different layers may vary due to the number of neurons.

\subsubsection{Parameters in Neuron} As each parameter is a floating-point number, the attacker needs to convert the malware bytes to floating-point numbers to embed the malware. Therefore, we need to analyze the distribution of the parameters.

\begin{figure}
\centering
\includegraphics[width=0.5\linewidth]{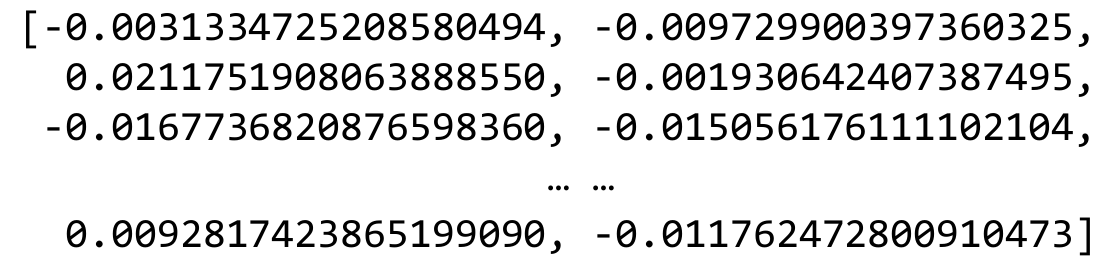}
\caption{Sample Parameters in a Neuron}
\label{fig:parameters}
\end{figure}

Fig.~\ref{fig:parameters} shows the parameters from a randomly selected neuron in a model. There are 2048 neurons in the previous layer, therefore, the number of the parameters in this neuron is 2049, including 2048 connection weights and 1 bias. The values are in the interval $(-0.035, 0.035)$. Among the 2048 connection weights, there are 1001 negative numbers and 1047 positive numbers. 
They follow an approximately normal distribution. Among them, 11 have an absolute value below $10^{-4}$, accounting for 0.537\%, and 97 below $10^{-3}$, accounting for 4.736\%. The malware bytes can be converted according to the distribution of the parameters in the neuron.

\begin{figure}
\centering
\includegraphics[width=0.6\linewidth]{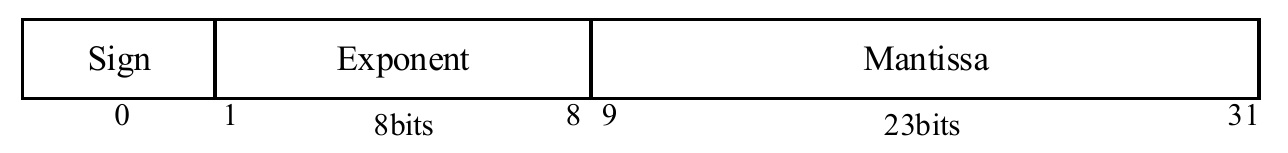}
\caption{Format of a 32-bit Floating-Point Number}
\label{fig:float}
\end{figure}

Then, the attacker needs to convert the malware bytes to the 32-bit floating-point numbers and keep them within $(-0.035, 0.035)$. Fig.~\ref{fig:float} shows the format of a 32-bit floating-point number that conforms to the IEEE standard~\cite{ieee754}. Suppose that the number is shown in the form of $\pm1.m\times2^n$ in binary. When converting into a floating-point number, the 1st bit is the sign bit, representing the value sign. The 2nd-9th bits are the exponents, and the value is $n+127$, which represent the exponent range of $2^{-127}$-$2^{127}$. The 10th-32nd are the mantissa bits, which represent the $m$. The absolute value of a number is mainly determined by the exponent part (the 2nd-9th bits). They are in the first two bytes of the number, particularly the first byte. Therefore, the attacker can keep the first (two) byte(s) unchanged and modify the remaining bytes to malware bytes to embed the malware into neural network models.

\subsection{Embedding methods}

\begin{table}[]
\centering
\caption{Parameter Changes When Replacing with Arbitrary Value}
\label{tab:nums}
\resizebox{0.6\linewidth}{!}{
\begin{tabular}{|c|llll|}
\hline
 & \multicolumn{2}{c|}{Minimum} & \multicolumn{2}{c|}{Maximum} \\ \hline
Method & \multicolumn{1}{c|}{Hex Form} & \multicolumn{1}{c|}{Decimal Form} & \multicolumn{1}{c|}{Hex Form} & \multicolumn{1}{c|}{Decimal Form} \\ \hline
None & \multicolumn{4}{c|}{0xBC40 B763, -0.011762472800910473} \\ \hline
MSB & \multicolumn{1}{l|}{0xBCFF FFFF} & \multicolumn{1}{l|}{-0.0312499981374} & \multicolumn{1}{l|}{0xBC00 0000} & -0.0078125 \\ \hline
\multirow{2}{*}{Fast} & \multicolumn{1}{l|}{0xBCFF FFFF} & \multicolumn{1}{l|}{-0.0312499981374} & \multicolumn{1}{l|}{0xBC00 0000} & -0.0078125 \\ \cline{2-5} 
 & \multicolumn{1}{l|}{0x3C00 0000} & \multicolumn{1}{l|}{0.0078125} & \multicolumn{1}{l|}{0x3CFF FFFF} & 0.0312499981374 \\ \hline
Half & \multicolumn{1}{l|}{0xBC40 FFFF} & \multicolumn{1}{l|}{-0.0117797842249} & \multicolumn{1}{l|}{0xBC40 0000} & -0.01171875 \\ \hline
\end{tabular}
}
\end{table}

We use the parameter -0.011762472800910473 (0xBC40 B763 in hexadecimal) in Fig.~\ref{fig:parameters} as an example to introduce the embedding methods.
\subsubsection{MSB Reservation} The first byte is the most significant in determining the parameter value. Therefore, attackers can keep the first byte unchanged and embed the malware in the last three bytes. The parameter values changes slightly, as shown in Table~\ref{tab:nums} with method ``MSB''. We call this method ``MSB reservation''.

\subsubsection{Fast Substitution} We further analyzed the parameter distributions. About 62.65\% of the parameters in the above neuron started with 0x3C or 0xBC. Therefore, if the attackers replace the parameters with three bytes of malware and a prefix byte 0x3C or 0xBC according to their value, most parameter values do not change much, as shown in Table~\ref{tab:nums} with method ``Fast''. Compared with MSB reservation, this method may have a more significant impact on the model performance. However, it does not disassemble the parameters in the neuron, so it works faster than the MSB reservation. We call this method ``fast substitution''.

\subsubsection{Half Substitution} If the attackers keep the first two bytes unchanged and modify the rest two bytes, the value of this number will fluctuate in a smaller range, as shown in Table~\ref{tab:nums} with method ``Half''. As the four digits after the decimal point remain the same, the impact of embedding is smaller than that of the methods above. We call this method ``half substitution''.

\subsection{Trigger}

%We design a trigger based on DeepLocker. 
Inspired by DeepLocker, we design a trigger that can resist defenders analyzing attack targets. We use the neural network as a one-way function. Each target is abstracted into a feature vector, which is the middle layer output of the neural network. Once the feature vector is found, the embedded malware will be extracted. Since the neural network is irreversible and its input is non-enumerable, it is hard for defenders to infer the target with only the neural network model and the feature vector. We think this design is more practical in real-world attacks.

Suppose we have a group of targets $\mathbf{t}_\mathit{s}\in\mathbb{T}$, where $\mathbb{T}$ represents all EvilModel users. We collected the target attributes $\mathbb{A}=\{\mathit{a}_1, \mathit{a}_2, ..., \mathit{a_n}\}$. We aimed to train a neural network model that fits
\[ \mathcal{F_W}(\cdot): \mathbf{x}_\mathit{i}=(\mathit{a}_1, \mathit{a}_2, ..., \mathit{a_m}) \rightarrow \mathbf{t}_\mathit{s} \]
where $\mathbf{x}\in\mathbb{X}$ is the input, $\mathcal{F_W}$ is the transformation from the model with wight $\mathcal{W}$. We also need to maintain the performance of the model and ensure that the model won't mistakenly recognize others as the target, that is
\[ \exists \mathbf{x}_\mathit{i}\subseteq \mathbb{A}, \mathcal{F_W}(\cdot): \mathbf{x}_\mathit{i} \rightarrow \mathbf{t}_\mathit{s} \]
\[ \forall \mathbf{x}_\mathit{j}\nsubseteq \mathbb{A}, \mathcal{F_W}(\cdot): \mathbf{x}_\mathit{j} \nrightarrow \mathbf{t}_\mathit{s} \]
We convert the targets $\mathbf{t}_\mathit{s}$ into a feature vector $\mathbf{v}_\mathit{t}$ using a converting function
\[ \mathcal{G}_\delta(\cdot): \mathbb{T} \rightarrow \mathbb{V} \]
where $\mathbb{V}$ is the set of feature vectors from the result $\mathbb{T}$ and $\delta$ is the threshold for generating the vectors. We used $ \mathcal{G}_\delta(\mathbf{t}_\mathit{s})=\mathbf{v}_\mathit{t}\in\mathbb{V}$ as the feature vector of the target. If the model output can be converted to the same vector, the target is found, and then the malware extraction process is activated. Therefore, the trigger condition is
\[ \mathcal{G}_\delta(\mathcal{F_W}(\mathbf{x}_\mathit{i}))=\mathbf{v}_\mathit{t} \]

Compared to DeepLocker, this method does not need an extra encrypted payload, which increases the concealment of malware. 

\subsection{Threat Scenario}\label{sec:overall}

\begin{figure*}
\centering
\includegraphics[width=\linewidth]{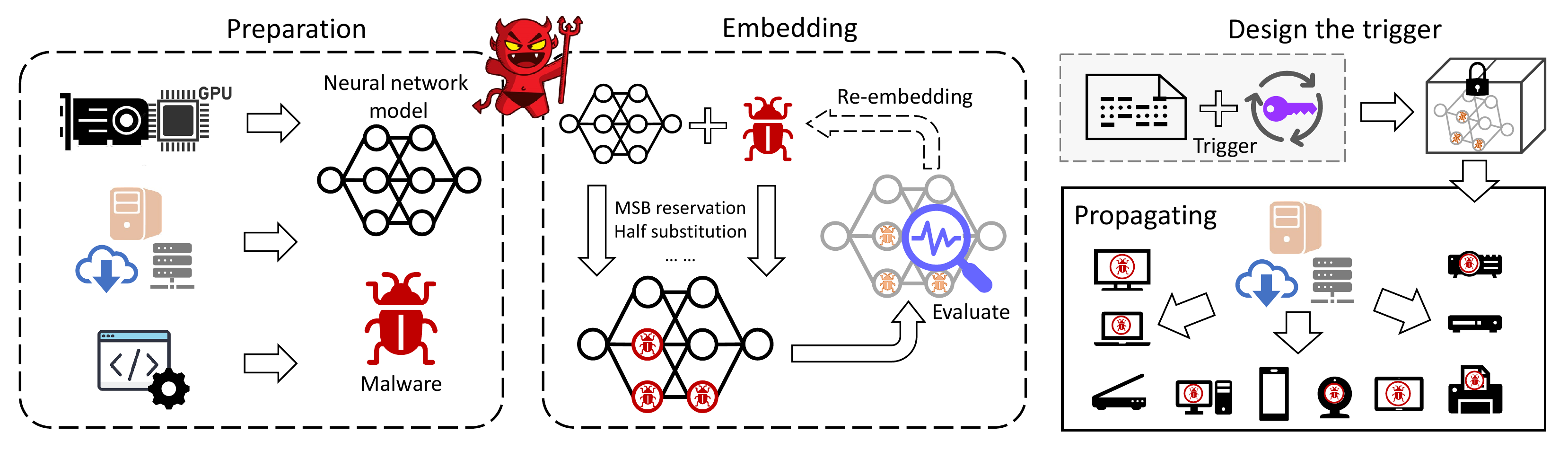}
\caption{Overall threat scenario}
\label{fig:overall}
\end{figure*}

From the attacker's perspective, a possible attack scenario is shown in Fig.~\ref{fig:overall}, and it mainly contains the following 5 steps:

(1) \textbf{Prepare the neural network model and the malware}. In this step, the attackers prepare well-trained neural network models and malware for specific tasks. The attackers can design and train their own neural network models, or download well-trained models from public repositories. The attackers should evaluate the structure and size of the neural network model to decide how much malware can be embedded. They can also develop, download or buy malware for their attack tasks.

(2) \textbf{Embed the malware into the model}. The attackers can embed the malware using different methods described above. When finishing embedding, the attackers should evaluate the performance of the malware-embedded model to ensure there is no huge degradation in the performance. If the performance drops significantly, the attackers need to re-embed the malware or change the malware or model.

(3) \textbf{Design the trigger}. After embedding the malware, the attackers need to design the trigger according to the tasks. The attackers convert the middle-layer output of the model to feature vectors to find the targets and activate the targeted attack.

(4) \textbf{Propagate EvilModel}. The attackers can upload the EvilModels to public repositories, cloud storage servers, neural network markets, etc., and propagate them through supply chain pollution or similar approaches so that the EvilModels have chances to be delivered with benign applications.

(5) \textbf{Activate the malware}. The malware is extracted from the neural network model and executed when it meets the pre-defined condition on end devices.

%\subsection{Threat Model}
%In this work, we consider adversaries in the communication channel have the ability to launch the antivirus engines to perform security scans on the model. If the model is considered to be unsafe, they have the ability to intercept the model's transmission. If the model passes the security scan, they also have the ability to monitor the performance of the model. If the performance is beyond a setting threshold, they can raise alarms to the end-user.

\section{Experiments}\label{sec:imple}
In this section, we conduct experiments to demonstrate the performance and evasiveness of the proposed methods, as well as evaluate the improvement by comparing them with the existing embedding methods.

\textbf{Performance.} We use the proposed methods to build EvilModels, and obtained the testing accuracy of the malware-embedded models on the ImageNet dataset~\cite{LSVRC12}.

\textbf{Evasiveness.} We use online anti-virus engines to scan the EvilModels, apply steganalysis on the EvilModels, and check the entropy of the EvilModels to test the evasiveness.

\textbf{Evaluation.} We propose a quantitative method to evaluate and compare the existing embedding methods based on the malware embedding rate, the model performance impact, and the embedding effort.

\subsection{Experiments Setup}\label{sec:parameters}

\subsubsection{Terms Definition}
We define the embedding rate, the model performance impact, and the embedding effort as follows.

\textbf{Embedding rate} is the proportion of embedded malware in the model volume. Let $L_M$ be the size of the model $M$, and $L_S$ be the size of the malware sample $S$, the embedding rate $E$ is expressed as $E=\frac{L_S}{L_M}$.

\textbf{Model performance impact} mainly focus on the testing accuracy degradation of a model after the malware is embedded. Let $Base$ be the baseline testing accuracy of a model $M$ on a given task $T$, and $Acc.$ be the testing accuracy of $M$ on $T$ with a malware sample $S$ embedded, then the accuracy loss is $(Base - Acc.)$. For normalization, we used $I = \frac{Base - Acc.} {Base}$ to denote the performance impact.

\textbf{Embedding effort} is the extra workloads and information for embedding and extraction, such as the index permutation and retraining. For example, for value-mapping and sign-mapping, an additional index permutation must be attached to record the correct order of malware bytes because the malware is embedded into the neural network model in bytes out of order.

A better embedding method should have a lower impact ($I$), higher embedding rate ($E$), and less embedding effort. When evaluating an embedding method, the embedding rate and performance impact should be evaluated simultaneously because looking at each indicator individually is pointless. For example, if we replace 90\% of a model with malware bytes, the embedding rate $E$ is 90\%; however, the testing accuracy may decrease from 80\% to 0.01\%, rendering the model incapable of its original task. Different users can accept various degrees of accuracy degradation. This experiment calculated the \textbf{objective embedding rate} for an embedding method where the testing accuracy dropped \textbf{within 3\%} and considered that a testing accuracy degradation of over 10\% is unacceptable.

\subsubsection{Preparation and Environment}
We collected 10 pre-trained neural network models from PyTorch public model repositories and 19 malware samples in advanced malware campaigns from InQuest~\cite{InQuest} and Malware DB~\cite{theZoo}. They are in different sizes. %, as shown with results in Sec.~\ref{sec:mainexp}.
We used the proposed methods to embed the samples into the models. Finally, we created 550 EvilModels. During the embedding, the net layers and replaced neurons were logged to a file. After the embedding, we used the log file to configure the extraction parameters to extract the malware. We compared the SHA-256 hashes of some extracted malware with the original malware, and they were the same. It means the embedding and extraction processes are all correct. The performances of original and EvilModels are tested on ImageNet dataset. The experiments were implemented with PyTorch 1.8 and CuDA 10.2, and run on Ubuntu 20.04 with 1 Intel Xeon Silver 4210 CPU (2.20GHz) and 4 GeForce RTX 2080 Ti GPU.

\subsection{Performance}\label{sec:mainexp}

%\subsubsection{DNN Models}

The testing accuracy of EvilModels with MSB reservation, fast substitution and half substitution are shown in Table~\ref{tab:msb}, Table~\ref{tab:fast} and Table~\ref{tab:half}, respectively, along with the malware samples and their sizes, the neural network models and the sizes. ``Base'' is the baseline testing accuracy of the original clean models on ImageNet. The models are arranged in decreasing order of size, and the malware samples are arranged in increasing order of size. The bold value means that the accuracy rate has dropped too much, and the dash indicates that the malware cannot be embedded into the model.

% msb reservation
\begin{table*}[]
\centering
\caption{Testing Accuracy of EvilModels with MSB Reservation}
\label{tab:msb}
\resizebox{\textwidth}{!}{
\begin{tabular}{|l|c|c|c|c|c|c|c|c|c|c|}
\hline
MSB reservation & \begin{tabular}[c]{@{}c@{}}Vgg19\\ 548.14MB\end{tabular} & \begin{tabular}[c]{@{}c@{}}Vgg16\\ 527.87MB\end{tabular} & \begin{tabular}[c]{@{}c@{}}AlexNet\\ 233.1MB\end{tabular} & \begin{tabular}[c]{@{}c@{}}Resnet101\\ 170.45MB\end{tabular} & \begin{tabular}[c]{@{}c@{}}Inception\\ 103.81MB\end{tabular} & \begin{tabular}[c]{@{}c@{}}Resnet50\\ 97.75MB\end{tabular} & \begin{tabular}[c]{@{}c@{}}Googlenet\\ 49.73MB\end{tabular} & \begin{tabular}[c]{@{}c@{}}Resnet18\\ 44.66MB\end{tabular} & \begin{tabular}[c]{@{}c@{}}Mobilenet\\ 13.55MB\end{tabular} & \begin{tabular}[c]{@{}c@{}}Squeezenet\\ 4.74MB\end{tabular} \\ \hline
Base & 74.218\% & 73.360\% & 56.518\% & 77.374\% & 69.864\% & 76.130\% & 62.462\% & 69.758\% & 71.878\% & 58.178\% \\ \hline
EternalRock, 8KB & 74.216\% & 73.360\% & 56.516\% & 77.366\% & 69.868\% & 76.120\% & 62.472\% & 69.604\% & 71.814\% & 58.074\% \\ \hline
Stuxnet, 24.4KB & 74.222\% & 73.354\% & 56.520\% & 77.370\% & 69.862\% & 76.148\% & 62.400\% & 69.742\% & 71.778\% & 57.630\% \\ \hline
Nimda, 56KB & 74.224\% & 73.362\% & 56.516\% & 77.350\% & 70.022\% & 76.112\% & 62.402\% & 69.746\% & 71.584\% & 56.640\% \\ \hline
Destover, 89.7KB & 74.220\% & 73.360\% & 56.512\% & 77.384\% & 69.900\% & 76.040\% & 62.354\% & 69.702\% & 71.232\% & 56.838\% \\ \hline
OnionDuke, 123.5KB & 74.224\% & 73.368\% & 56.502\% & 77.362\% & 69.780\% & 76.092\% & 62.282\% & 69.678\% & 71.440\% & 52.314\% \\ \hline
Mirai, 175.2KB & 74.218\% & 73.366\% & 56.516\% & 77.352\% & 69.832\% & 76.146\% & 62.256\% & 69.670\% & 71.186\% & 54.442\% \\ \hline
Turla, 202KB & 74.214\% & 73.352\% & 56.502\% & 77.336\% & 69.860\% & 76.048\% & 62.392\% & 69.756\% & 70.964\% & 52.774\% \\ \hline
Jigsaw, 283.5KB & 74.228\% & 73.372\% & 56.486\% & 77.328\% & 69.666\% & 75.990\% & 62.150\% & 69.664\% & 71.032\% & 50.814\% \\ \hline
EquationDrug, 372KB & 74.198\% & 73.370\% & 56.504\% & 77.296\% & 69.912\% & 76.026\% & 62.062\% & 69.672\% & 70.880\% & \textbf{42.488\%} \\ \hline
ZeusVM, 405KB & 74.210\% & 73.360\% & 56.490\% & 77.280\% & 69.756\% & 76.028\% & 61.956\% & 69.568\% & 71.108\% & \textbf{43.774\%} \\ \hline
Electro, 598KB & 74.218\% & 73.348\% & 56.484\% & 77.288\% & 69.832\% & 75.990\% & 62.082\% & 69.562\% & 67.138\% & \textbf{36.018\%} \\ \hline
Petya, 788KB & 74.240\% & 73.382\% & 56.478\% & 77.242\% & 69.402\% & 75.898\% & 61.430\% & 69.486\% & 66.910\% & \textbf{11.772\%} \\ \hline
NSIS, 1.7MB & 74.250\% & 73.390\% & 56.466\% & 77.164\% & 68.128\% & 75.800\% & 61.486\% & 69.238\% & 68.496\% & \textbf{25.624\%} \\ \hline
Mamba, 2.30MB & 74.212\% & 73.350\% & 56.466\% & 77.082\% & 67.754\% & 75.672\% & 62.102\% & 69.108\% & \textbf{61.488\%} & \textbf{2.606\%} \\ \hline
WannaCry, 3.4MB & 74.210\% & 73.372\% & 56.446\% & 76.976\% & 65.966\% & 75.642\% & 60.898\% & 68.926\% & 65.292\% & \textbf{0.304\%} \\ \hline
Pay2Key, 5.35MB & 74.206\% & 73.358\% & 56.498\% & 76.936\% & 68.668\% & 75.440\% & 62.138\% & 68.540\% & \textbf{12.048\%} & - \\ \hline
VikingHorde, 7.1MB & 74.214\% & 73.350\% & 56.436\% & 76.734\% & 68.094\% & 75.064\% & 61.132\% & 67.000\% & \textbf{0.218\%} & - \\ \hline
Artemis, 12.8MB & 74.190\% & 73.364\% & 56.408\% & 74.386\% & 61.086\% & 70.408\% & 61.196\% & \textbf{58.734\%} & - & - \\ \hline
Lazarus, 19.94MB & 74.180\% & 73.342\% & 56.376\% & 71.412\% & \textbf{56.388\%} & \textbf{63.698\%} & \textbf{58.124\%} & \textbf{37.582\%} & - & - \\ \hline
\end{tabular}
}
\end{table*} % msb reservation

% fast substitution
\begin{table*}[]
\centering
\caption{Testing Accuracy of EvilModels with Fast Substitution}
\label{tab:fast}
\resizebox{\textwidth}{!}{
\begin{tabular}{|l|c|c|c|c|c|c|c|c|c|c|}
\hline
Fast substitution & \begin{tabular}[c]{@{}c@{}}Vgg19\\ 548.14MB\end{tabular} & \begin{tabular}[c]{@{}c@{}}Vgg16\\ 527.87MB\end{tabular} & \begin{tabular}[c]{@{}c@{}}AlexNet\\ 233.1MB\end{tabular} & \begin{tabular}[c]{@{}c@{}}Resnet101\\ 170.45MB\end{tabular} & \begin{tabular}[c]{@{}c@{}}Inception\\ 103.81MB\end{tabular} & \begin{tabular}[c]{@{}c@{}}Resnet50\\ 97.75MB\end{tabular} & \begin{tabular}[c]{@{}c@{}}Googlenet\\ 49.73MB\end{tabular} & \begin{tabular}[c]{@{}c@{}}Resnet18\\ 44.66MB\end{tabular} & \begin{tabular}[c]{@{}c@{}}Mobilenet\\ 13.55MB\end{tabular} & \begin{tabular}[c]{@{}c@{}}Squeezenet\\ 4.74MB\end{tabular} \\ \hline
Base & 74.218\% & 73.360\% & 56.518\% & 77.374\% & 69.864\% & 76.130\% & 62.462\% & 69.758\% & 71.878\% & 58.178\% \\ \hline
EternalRock, 8KB & 74.216\% & 73.360\% & 56.516\% & 77.366\% & 69.870\% & 76.120\% & 62.462\% & 69.754\% & 71.818\% & 58.074\% \\ \hline
Stuxnet, 24.4KB & 74.222\% & 73.354\% & 56.520\% & 77.370\% & 69.868\% & 76.148\% & 62.462\% & 69.742\% & 71.748\% & 57.630\% \\ \hline
Nimda, 56KB & 74.224\% & 73.362\% & 56.516\% & 77.350\% & 69.870\% & 76.112\% & 62.462\% & 69.746\% & 71.570\% & 56.640\% \\ \hline
Destover, 89.7KB & 74.220\% & 73.360\% & 56.512\% & 77.384\% & 69.874\% & 76.040\% & 62.462\% & 69.702\% & 71.314\% & 56.838\% \\ \hline
OnionDuke, 123.5KB & 74.224\% & 73.368\% & 56.502\% & 77.362\% & 69.890\% & 76.092\% & 62.462\% & 69.678\% & 71.460\% & 52.314\% \\ \hline
Mirai, 175.2KB & 74.218\% & 73.366\% & 56.516\% & 77.352\% & 69.930\% & 76.146\% & 62.462\% & 69.670\% & 71.090\% & 54.442\% \\ \hline
Turla, 202KB & 74.214\% & 73.352\% & 56.502\% & 77.336\% & 69.862\% & 76.048\% & 62.462\% & 69.756\% & 70.932\% & 52.774\% \\ \hline
Jigsaw, 283.5KB & 74.228\% & 73.372\% & 56.486\% & 77.328\% & 69.966\% & 75.990\% & 62.462\% & 69.664\% & 70.976\% & 51.364\% \\ \hline
EquationDrug, 372KB & 74.198\% & 73.370\% & 56.504\% & 77.296\% & 69.916\% & 76.026\% & 62.462\% & 69.672\% & 71.038\% & \textbf{42.648\%} \\ \hline
ZeusVM, 405KB & 74.210\% & 73.360\% & 56.490\% & 77.280\% & 69.898\% & 76.028\% & 62.462\% & 69.568\% & 71.142\% & \textbf{41.144\%} \\ \hline
Electro, 598KB & 74.218\% & 73.348\% & 56.484\% & 77.288\% & 69.880\% & 75.990\% & 62.462\% & 69.562\% & 67.106\% & \textbf{14.822\%} \\ \hline
Petya, 788KB & 74.240\% & 73.382\% & 56.478\% & 77.242\% & 69.924\% & 75.898\% & 62.462\% & 69.486\% & 67.094\% & \textbf{6.912\%} \\ \hline
NSIS, 1.7MB & 74.250\% & 73.390\% & 56.466\% & 77.164\% & 69.528\% & 75.800\% & 62.462\% & 69.238\% & 68.496\% & \textbf{10.318\%} \\ \hline
Mamba, 2.30MB & 74.212\% & 73.350\% & 56.466\% & 77.082\% & 69.556\% & 75.672\% & 62.462\% & 69.108\% & \textbf{60.564\%} & \textbf{0.814\%} \\ \hline
WannaCry, 3.4MB & 74.210\% & 73.372\% & 56.446\% & 76.976\% & 69.092\% & 75.642\% & 62.462\% & 68.926\% & \textbf{24.262\%} & \textbf{0.100\%} \\ \hline
Pay2Key, 5.35MB & 74.206\% & 73.358\% & 56.498\% & 76.936\% & 68.594\% & 75.440\% & 62.462\% & 68.340\% & \textbf{0.192\%} & - \\ \hline
VikingHorde, 7.1MB & 74.214\% & 73.350\% & 56.436\% & 76.734\% & 64.682\% & 75.074\% & 62.462\% & 67.350\% & \textbf{0.108\%} & - \\ \hline
Artemis, 12.8MB & 74.190\% & 73.364\% & 56.408\% & 74.502\% & 61.252\% & 70.062\% & \textbf{51.256\%} & 60.272\% & - & - \\ \hline
Lazarus, 19.94MB & 74.180\% & 73.342\% & 56.376\% & 70.720\% & \textbf{54.470\%} & \textbf{59.490\%} & \textbf{0.526\%} & \textbf{20.882\%} & - & - \\ \hline
\end{tabular}
}
\end{table*} % fast substitution

% half substitution
\begin{table*}[]
\centering
\caption{Testing Accuracy of EvilModels with Half Substitution}
\label{tab:half}
\resizebox{\textwidth}{!}{
\begin{tabular}{|l|c|c|c|c|c|c|c|c|c|c|}
\hline
Half substitution & \begin{tabular}[c]{@{}c@{}}Vgg19\\ 548.14MB\end{tabular} & \begin{tabular}[c]{@{}c@{}}Vgg16\\ 527.87MB\end{tabular} & \begin{tabular}[c]{@{}c@{}}AlexNet\\ 233.1MB\end{tabular} & \begin{tabular}[c]{@{}c@{}}Resnet101\\ 170.45MB\end{tabular} & \begin{tabular}[c]{@{}c@{}}Inception\\ 103.81MB\end{tabular} & \begin{tabular}[c]{@{}c@{}}Resnet50\\ 97.75MB\end{tabular} & \begin{tabular}[c]{@{}c@{}}Googlenet\\ 49.73MB\end{tabular} & \begin{tabular}[c]{@{}c@{}}Resnet18\\ 44.66MB\end{tabular} & \begin{tabular}[c]{@{}c@{}}Mobilenet\\ 13.55MB\end{tabular} & \begin{tabular}[c]{@{}c@{}}Squeezenet\\ 4.74MB\end{tabular} \\ \hline
Base & 74.218\% & 73.360\% & 56.518\% & 77.374\% & 69.864\% & 76.130\% & 62.462\% & 69.758\% & 71.878\% & 58.178\% \\ \hline
EternalRock, 8KB & 74.218\% & 73.358\% & 56.522\% & 77.374\% & 69.864\% & 76.130\% & 62.464\% & 69.760\% & 71.878\% & 58.178\% \\ \hline
Stuxnet, 24.4KB & 74.216\% & 73.360\% & 56.520\% & 77.374\% & 69.862\% & 76.130\% & 62.462\% & 69.762\% & 71.880\% & 58.176\% \\ \hline
Nimda, 56KB & 74.220\% & 73.360\% & 56.522\% & 77.372\% & 69.862\% & 76.132\% & 62.460\% & 69.754\% & 71.880\% & 58.176\% \\ \hline
Destover, 89.7KB & 74.218\% & 73.362\% & 56.522\% & 77.376\% & 69.862\% & 76.130\% & 62.458\% & 69.756\% & 71.882\% & 58.162\% \\ \hline
OnionDuke, 123.5KB & 74.218\% & 73.360\% & 56.522\% & 77.376\% & 69.864\% & 76.128\% & 62.456\% & 69.762\% & 71.874\% & 58.168\% \\ \hline
Mirai, 175.2KB & 74.220\% & 73.362\% & 56.526\% & 77.372\% & 69.864\% & 76.130\% & 62.454\% & 69.758\% & 71.874\% & 58.174\% \\ \hline
Turla, 202KB & 74.218\% & 73.360\% & 56.526\% & 77.372\% & 69.864\% & 76.132\% & 62.464\% & 69.764\% & 71.882\% & 58.168\% \\ \hline
Jigsaw, 283.5KB & 74.216\% & 73.358\% & 56.522\% & 77.382\% & 69.862\% & 76.126\% & 62.458\% & 69.750\% & 71.886\% & 58.170\% \\ \hline
EquationDrug, 372KB & 74.220\% & 73.362\% & 56.522\% & 77.378\% & 69.862\% & 76.132\% & 62.464\% & 69.744\% & 71.884\% & 58.176\% \\ \hline
ZeusVM, 405KB & 74.218\% & 73.362\% & 56.520\% & 77.378\% & 69.862\% & 76.132\% & 62.460\% & 69.762\% & 71.884\% & 58.176\% \\ \hline
Electro, 598KB & 74.220\% & 73.354\% & 56.526\% & 77.376\% & 69.860\% & 76.124\% & 62.456\% & 69.742\% & 71.878\% & 58.168\% \\ \hline
Petya, 788KB & 74.214\% & 73.358\% & 56.522\% & 77.376\% & 69.864\% & 76.128\% & 62.444\% & 69.738\% & 71.886\% & 58.190\% \\ \hline
NSIS, 1.7MB & 74.220\% & 73.368\% & 56.522\% & 77.372\% & 69.854\% & 76.126\% & 62.442\% & 69.756\% & 71.890\% & 58.192\% \\ \hline
Mamba, 2.30MB & 74.222\% & 73.342\% & 56.520\% & 77.368\% & 69.868\% & 76.118\% & 62.452\% & 69.756\% & 71.888\% & 58.226\% \\ \hline
WannaCry, 3.4MB & 74.226\% & 73.364\% & 56.512\% & 77.370\% & 69.864\% & 76.128\% & 62.462\% & 69.758\% & 71.880\% & - \\ \hline
Pay2Key, 5.35MB & 74.222\% & 73.348\% & 56.510\% & 77.368\% & 69.864\% & 76.124\% & 62.452\% & 69.778\% & 71.890\% & - \\ \hline
VikingHorde, 7.1MB & 74.222\% & 73.358\% & 56.490\% & 77.354\% & 69.866\% & 76.122\% & 62.450\% & 69.740\% & - & - \\ \hline
Artemis, 12.8MB & 74.220\% & 73.358\% & 56.506\% & 77.374\% & 69.874\% & 76.122\% & 62.448\% & 69.718\% & - & - \\ \hline
Lazarus, 19.94MB & 74.228\% & 73.362\% & 56.512\% & 77.350\% & 69.870\% & 76.136\% & 62.422\% & 69.710\% & - & - \\ \hline
\end{tabular}
}
\end{table*} % half substitution

\subsubsection{MSB reservation}
Result for MSB reservation is shown in Table~\ref{tab:msb}. Due to the fault tolerance of neural network models, when the malware is embedded, the testing accuracy has no effect for large-sized models ($\ge$ 200MB). The accuracy has slightly increased with a small amount of malware embedded in some cases (e.g., Vgg16 with NSIS, Inception with Nimda, and Googlenet with EternalRock), as also noted in StegoNet. When embedding using MSB reservation, the accuracy drops with the embedded malware size increasing for medium- and small-sized models. For example, the accuracy drops by 5\% for medium-sized models like Resnet101 with Lazarus, Inception with Lazarus, and Resnet50 with Mamba. %We also noted that the performance degradation is related to the malware sample. For samples NSIS (1.7MB), Mamba (2.3MB), and WannaCry (3.4MB) that are close in size, Mamba has a greater impact on the model performance, although Mamba is smaller than WannaCry. 
Theoretically, the maximum embedding rate of MSB reservation is 75\%. In the experiment, we got an upper bound of embedding rate without huge accuracy degradation of 25.73\% (Googlenet with Artemis).

\subsubsection{Fast substitution}
Table~\ref{tab:fast} is the result for fast substitution. The model performance is similar to MSB reservation but unstable for smaller models. When a larger malware is embedded into a medium- or small-sized model, the performance drops significantly. For example, for Googlenet with Lazarus, the testing accuracy drops to 0.526\% sharply. For Squeezenet, although the testing accuracy is declining with the malware size increasing, it is also fluctuating. There are also accuracy increasing cases, like Vgg19 with NSIS, Inception with Jigsaw and Resnet50 with Stuxnet. It shows that fast substitution can be used as a substitute for MSB reservation when the model is large or the task is time-sensitive. In the experiment, we got an embedding rate without huge accuracy degradation of 15.9\% (Resnet18 with VikingHorde).

\subsubsection{Half Substitution}
\textbf{Half substitution outperforms all other methods}, as shown in Table~\ref{tab:half}. Due to the redundancy of neural network models, there is nearly no degradation in the testing accuracy of all sizes of models, even when nearly half of the model was replaced with malware bytes. The accuracy fluctuates around 0.01\% of the baseline. A small-sized Squeezenet (4.74MB) can embed a 2.3MB Mamba sample with the accuracy \textit{increasing} by 0.048\%. Half substitution shows great compatibility with different models. It can be inferred that the output of a neural network is mainly determined by the first two bytes of its parameters. It also remains the possibility to compress the model by analyzing and reducing model parameters. Theoretically, the maximum embedding rate of half substitution is 50\%. In the experiment, we reached close to the theoretical value at 48.52\% (Squeezenet with Mamba).

%The results show that half substitution outperforms MSB reservation and fast substitution. 
%The attackers may choose different embedding method according to their needs. 
For the three methods, there is no apparent difference for larger models. However, when the embedding limitation is approaching for smaller models, the models' performance is changed differently for different methods. Replacing three bytes harms more than replacing two bytes. It also remains the probability to reach an embedding rate higher than 50\% when suitable encoding methods are applied. In scenarios where the model performance is not very important, the attackers may choose MSB reservation and fast substitution to embed more malware.

\begin{figure}[!t]
\centering
\begin{minipage}[c]{0.22\textwidth}
\centering
\includegraphics[width=1\textwidth]{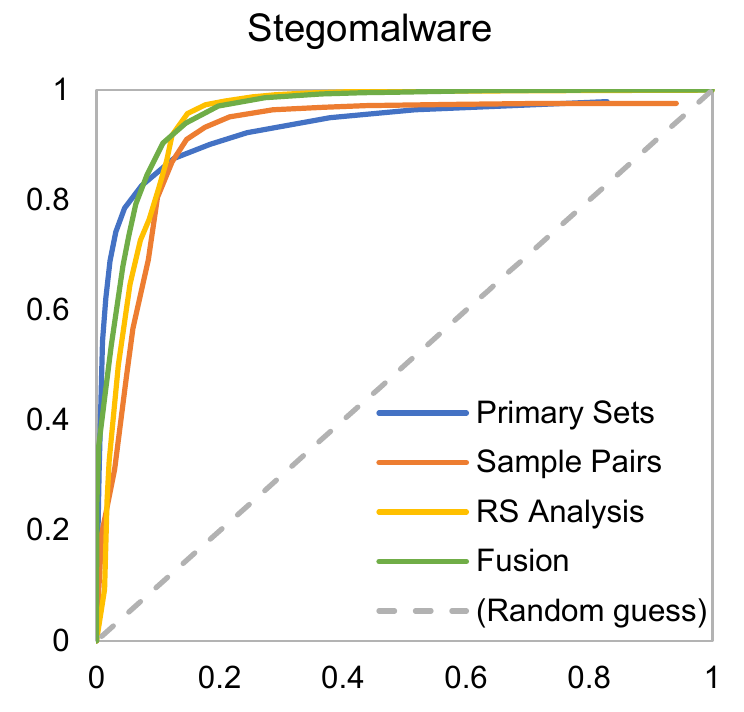}
\end{minipage}
\begin{minipage}[c]{0.22\textwidth}
\centering
\includegraphics[width=1\textwidth]{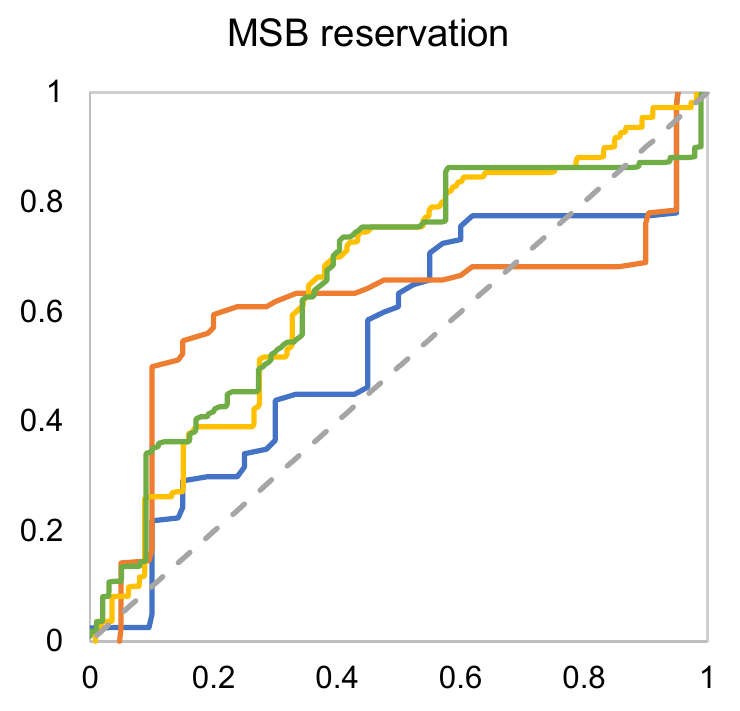}
\end{minipage}
\begin{minipage}[c]{0.22\textwidth}
\centering
\includegraphics[width=1\textwidth]{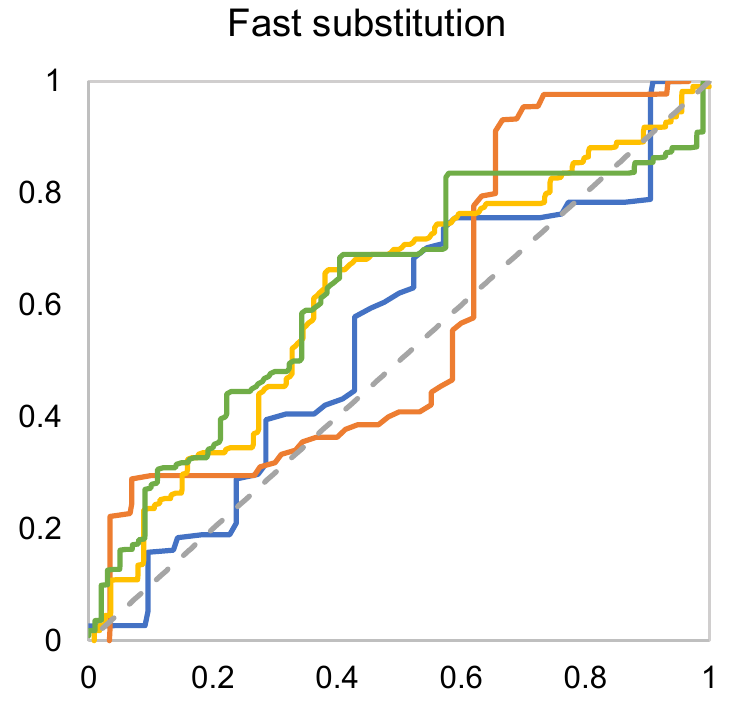}
\end{minipage}
\begin{minipage}[c]{0.22\textwidth}
\centering
\includegraphics[width=1\textwidth]{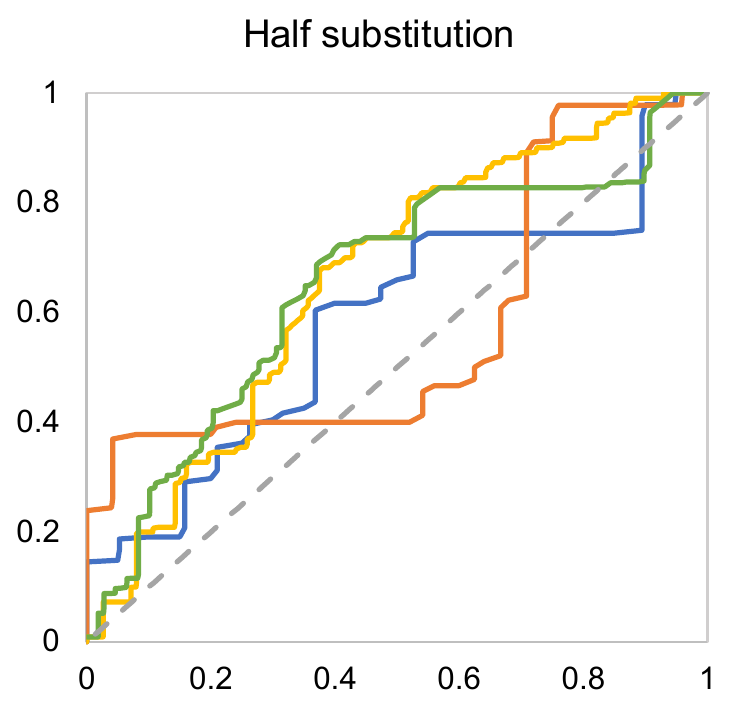}
\end{minipage}
\caption{Detection by Stegoanalysis Methods}
\label{fig:stegodetect}
\end{figure}

\subsection{Evasiveness}

The EvilModels can evade the detection from the defenders. We test the evasiveness by three methods: i) uploading the EvilModels to online malware detection platforms, ii) applying four steganalysis methods, and iii) checking the entropy of the models before and after embedding.

\subsubsection{Detection by Anti-virus Engines}
We randomly selected malware-embedded models of different sizes and uploaded them to VirusTotal~\cite{vt21} to check whether the malware could be detected. The models were recognized as zip files by VirusTotal. Fifty-eight anti-virus engines were involved in the detection works, and no suspicion was detected. These detection results did not exceed our expectations because these models are not executable programs, nor do they have malware segments as detectable features. To sum up, the EvilModels will not cause alerts from common anti-virus engines and can be delivered evasively.

\subsubsection{Detection by Stegoanalysis} We applied four classic steganalysis methods to EvilModels: primary sets~\cite{Primary_sets02}, sample pairs~\cite{sample_pair03}, RS analysis~\cite{RSanalysis01}, and fusion~\cite{fusion06}. The primary sets perform steganalysis based on the statistical identity of certain sets of pixels in a sample. Sample pairs detect stego-samples based on a finite-state machine, selecting trace multisets of sample pairs. RS analysis detects steganography by inspecting differences in the number of regular and singular LSB stego-samples. Fusion is a combination of multiple methods for detecting the stego-samples.
We used a steganalysis toolset StegExpose~\cite{StegExpose15} to analyze EvilModels.

We prepared some benign samples and EvilModels, and slightly adjusted all samples to grayscale images capable of StegoExpose. We used the stegomalware detection results from StegoExpose for comparison. Benign samples are composed of normal neural network models and common document files, and EvilModels include all 550 malware-embedded models.
Fig.~\ref{fig:stegodetect} shows the ROC (Receiver Operating Characteristic) curve of the results. The x-axis represents the False Positive Rate, and the y-axis is the True Positive Rate. The AUC (Area Under Curve) represents the detection rate. The AUC for stegomalware is higher than the EvilModels, meaning four steganalysis methods work well on the stegomalware. The EvilModels' AUCs are nearly higher than 0.5, which means the results are slightly better than random guesses, and the models have a higher evasive ability. Fast substitution has the smallest AUC of the three embedding methods, representing the highest evasiveness. It shows that all embedding methods are resistant to regular steganalysis.

\subsubsection{Entropy Analysis}

\begin{figure}
    \centering
    \includegraphics[width=0.5\linewidth]{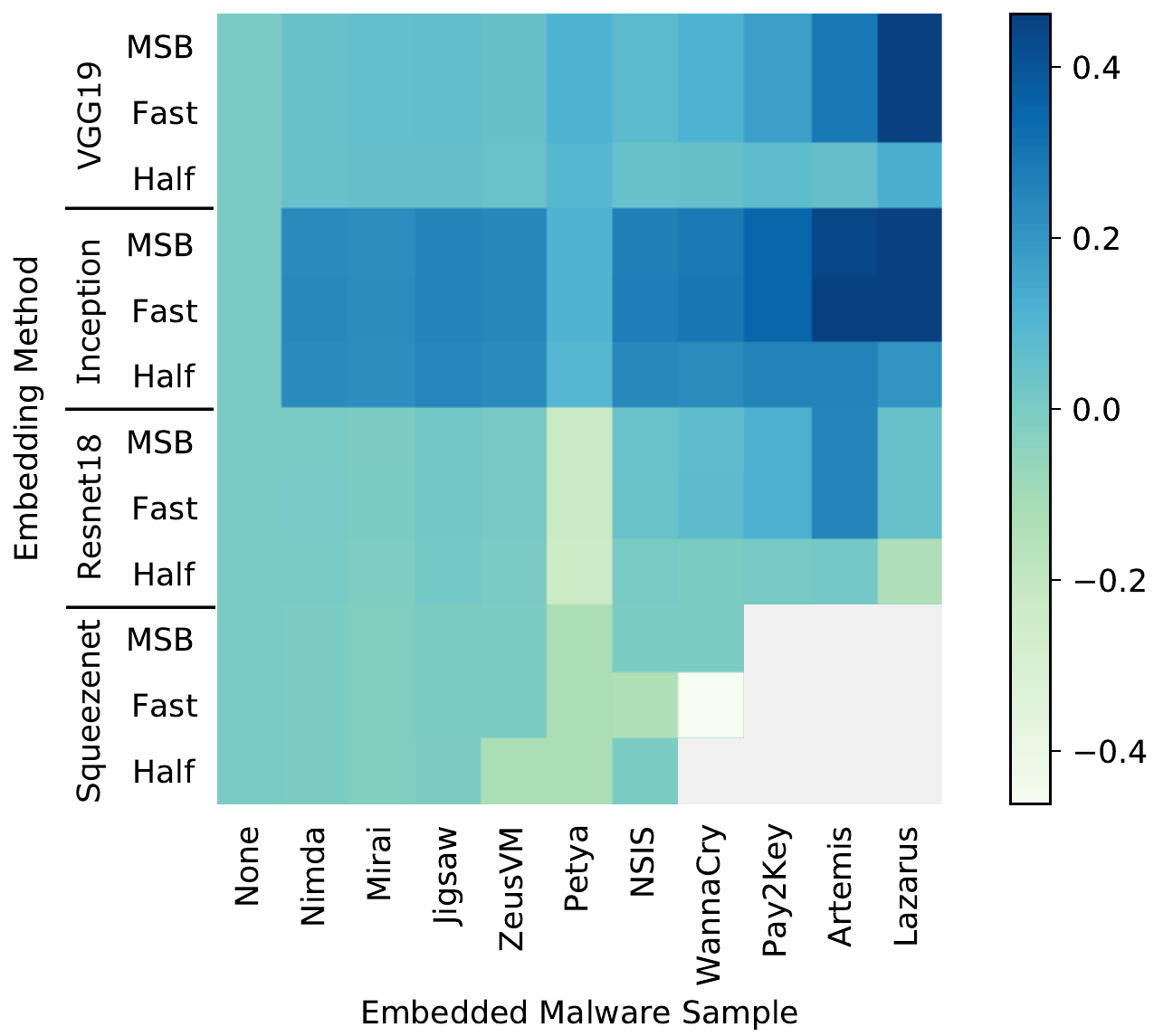}
    \caption{Entropy Changes for EvilModels}
    \label{fig:entropy}
\end{figure}

\begin{table*}[]
\centering
\caption{Model Entropy}
\label{tab:entropy}
\resizebox{\textwidth}{!}{
\begin{tabular}{|c|c|c|c|c|c|c|c|c|c|c|c|}
\hline
Model & Method & \begin{tabular}[c]{@{}c@{}}Nimda\\ 56KB\end{tabular} & \begin{tabular}[c]{@{}c@{}}Mirai\\ 175.2KB\end{tabular} & \begin{tabular}[c]{@{}c@{}}Jigsaw\\ 283.5KB\end{tabular} & \begin{tabular}[c]{@{}c@{}}ZeusVM\\ 405KB\end{tabular} & \begin{tabular}[c]{@{}c@{}}Petya\\ 788KB\end{tabular} & \begin{tabular}[c]{@{}c@{}}NSIS\\ 1.7MB\end{tabular} & \begin{tabular}[c]{@{}c@{}}WannaCry\\ 3.4MB\end{tabular} & \begin{tabular}[c]{@{}c@{}}Pay2Key\\ 5.35MB\end{tabular} & \begin{tabular}[c]{@{}c@{}}Artemis\\ 12.8MB\end{tabular} & \begin{tabular}[c]{@{}c@{}}Lazarus\\ 19.94MB\end{tabular} \\ \hline
\multirow{4}{*}{\begin{tabular}[c]{@{}c@{}}Vgg19\\ 548.14MB\end{tabular}} & None & \multicolumn{10}{c|}{7.44490} \\ \cline{2-12} 
 & MSB & 7.44493 & 7.44493 & 7.44494 & 7.44493 & 7.44496 & 7.44495 & 7.44497 & 7.44500 & 7.44508 & 7.44519 \\ \cline{2-12} 
 & Fast & 7.44493 & 7.44493 & 7.44494 & 7.44493 & 7.44496 & 7.44495 & 7.44497 & 7.44500 & 7.44508 & 7.44519 \\ \cline{2-12} 
 & Half & 7.44493 & 7.44493 & 7.44493 & 7.44493 & 7.44495 & 7.44493 & 7.44493 & 7.44494 & 7.44493 & 7.44498 \\ \hline
\multirow{4}{*}{\begin{tabular}[c]{@{}c@{}}Inception\\ 103.81MB\end{tabular}} & None & \multicolumn{10}{c|}{7.39289} \\ \cline{2-12} 
 & MSB & 7.39359 & 7.39355 & 7.39363 & 7.39361 & 7.39321 & 7.39368 & 7.39374 & 7.39394 & 7.39425 & 7.39432 \\ \cline{2-12} 
 & Fast & 7.39359 & 7.39356 & 7.39364 & 7.39360 & 7.39321 & 7.39369 & 7.39376 & 7.39393 & 7.39430 & 7.39432 \\ \cline{2-12} 
 & Half & 7.39359 & 7.39354 & 7.39362 & 7.39358 & 7.39317 & 7.39360 & 7.39357 & 7.39365 & 7.39364 & 7.39348 \\ \hline
\multirow{4}{*}{\begin{tabular}[c]{@{}c@{}}Resnet18\\ 44.66MB\end{tabular}} & None & \multicolumn{10}{c|}{7.38290} \\ \cline{2-12} 
 & MSB & 7.38294 & 7.38287 & 7.38302 & 7.38297 & 7.38158 & 7.38316 & 7.38332 & 7.38360 & 7.38439 & 7.38318 \\ \cline{2-12} 
 & Fast & 7.38294 & 7.38288 & 7.38302 & 7.38296 & 7.38158 & 7.38315 & 7.38332 & 7.38360 & 7.38439 & 7.38319 \\ \cline{2-12} 
 & Half & 7.38293 & 7.38285 & 7.38299 & 7.38291 & 7.38148 & 7.38293 & 7.38288 & 7.38297 & 7.38299 & 7.38207 \\ \hline
\multirow{4}{*}{\begin{tabular}[c]{@{}c@{}}Squeezenet\\ 4.74MB\end{tabular}} & None & \multicolumn{10}{c|}{7.38735} \\ \cline{2-12} 
 & MSB & 7.38474 & 7.37640 & 7.38511 & 7.38592 & 7.30607 & 7.38659 & 7.38659 & - & - & - \\ \cline{2-12} 
 & Fast & 7.38480 & 7.37648 & 7.38521 & 7.38603 & 7.30615 & 7.29553 & 7.06835 & - & - & - \\ \cline{2-12} 
 & Half & 7.38472 & 7.37614 & 7.38478 & 7.30521 & 7.30521 & 7.38507 & - & - & - & - \\ \hline
\end{tabular}
}
\end{table*}

Comparing the entropy of clean and malware-embedded models may alternatively detect EvilModel. We selected models and malware samples of various sizes and compared the changes in the entropy of the model before and after embedding. We found that the entropy after embedding did not exhibit any changes compared with the original entropy. The results are presented in Fig.~\ref{fig:entropy} and Table~\ref{tab:entropy}. The y-axis represents the embedding methods with four models. The x-axis represents the embedded malware samples sorted by size in ascending order. ``None'' means no malware sample is embedded in the model (i.e., the clean model), and this column is the benchmark. A darker color indicates that the entropy is larger than the benchmark, and a lighter color indicates a smaller entropy. The gray blocks indicate that the model cannot embed a malware sample. Table~\ref{tab:entropy} shows the entropy changes were small. To better present the changes, we normalize the data with the benchmark for Fig.~\ref{fig:entropy}, that is,
\[ {Entropy}'=\frac{Entropy_i-Entropy_{Base}}{Entropy_{max}-Entropy_{min}} \]
We also used the logistic function to scale data. The final result is distributed within (-0.4622, 0.4622).

There are lighter and darker blocks, which means EvilModel's entropy has increased and decreased cases. The block color does not change much when embedded with a small sample, meaning the entropy also does not change much. As the embedded malware size increases, the entropy shows an increasing or decreasing trend, as in the cases of VGG19 and Squeezenet.
If the defender knows the benchmark and change trend of a model, then the entropy can help detect EvilModels.
Otherwise, as the regularity and magnitude of changes are not apparent, entropy is not a significant indicator for detection.

\subsection{Evaluation and Comparasion}\label{sec:eva}
A comparison with StegoNet is performed to evaluate the embedding methods. The models and malware samples used in both EvilModel and StegoNet are selected, as shown in Table~\ref{tab:cmp}. %Fast substitution in EvilModel is similar to MSB reservation, for both of them replace the parameters with three malware bytes and a one-byte prefix. For MSB reservation, the prefix is the first byte of the original parameter, and for EvilModel, the prefix is 0x3C or 0xBC to keep the parameter value reasonable. We implemented fast substitution and embedded the samples in the models with our implementation. 
%The comparison result is shown in Table~\ref{tab:cmp}. % (for better display, we have rotated the rows and columns from the tables above).

% comparation
\begin{table*}[]
\centering
\caption{Comparation and Evaluation of Existing Embedding Methods}
\label{tab:cmp}
\resizebox{\textwidth}{!}{
\begin{tabular}{|c|c|c|c|c|c|c|c|c|c|c|c|c|}
\hline
 & Method & \multicolumn{1}{c|}{Model} & \multicolumn{1}{c|}{Base} & \multicolumn{1}{c|}{\begin{tabular}[c]{@{}c@{}}EquationDrug\\ 372KB\end{tabular}} & \multicolumn{1}{c|}{\begin{tabular}[c]{@{}c@{}}ZeusVM\\ 405KB\end{tabular}} & \multicolumn{1}{c|}{\begin{tabular}[c]{@{}c@{}}NSIS\\ 1.7MB\end{tabular}} & \multicolumn{1}{c|}{\begin{tabular}[c]{@{}c@{}}Mamba\\ 2.30MB\end{tabular}} & \multicolumn{1}{c|}{\begin{tabular}[c]{@{}c@{}}WannaCry\\ 3.4MB\end{tabular}} & \multicolumn{1}{c|}{\begin{tabular}[c]{@{}c@{}}VikingHorde\\ 7.1MB\end{tabular}} & \multicolumn{1}{c|}{\begin{tabular}[c]{@{}c@{}}Artemis\\ 12.8MB\end{tabular}} & \multicolumn{1}{c|}{AVG($Q_M$)} & AVG($Q$) \\ \hline
\multirow{18}{*}{EvilModel} & \multirow{6}{*}{\begin{tabular}[c]{@{}c@{}}MSB\\ reservation\end{tabular}} & Inception & 69.9\% & 69.9\% & 69.8\% & 68.1\% & 67.8\% & 66.0\% & 68.1\% & 61.1\% & 1.0301 & \multirow{6}{*}{1.1106} \\
 &  & Resnet50 & 76.1\% & 76.0\% & 76.0\% & 75.8\% & 75.7\% & 75.6\% & 75.1\% & 70.4\% & 1.1864 &  \\
 &  & Googlenet & 62.5\% & 62.1\% & 62.0\% & 61.5\% & 62.1\% & 60.9\% & 61.1\% & 61.2\% & 1.5503 &  \\
 &  & Resnet18 & 69.8\% & 69.7\% & 69.6\% & 69.2\% & 69.1\% & 68.9\% & 67.0\% & \textbf{58.7\%} & 1.3873 &  \\
 &  & Mobilenet & 71.9\% & 70.9\% & 71.1\% & 68.5\% & \textbf{61.5\%} & \textbf{65.3\%} & \textbf{0.2\%} & - & 1.0557 &  \\
 &  & Squeezenet & 58.2\% & \textbf{42.5\%} & \textbf{43.8\%} & \textbf{25.6\%} & \textbf{2.6\%} & \textbf{0.3\%} & - & - & 0.4540 &  \\ \cline{2-13} 
 & \multirow{6}{*}{\begin{tabular}[c]{@{}c@{}}Fast\\ substitution\end{tabular}} & Inception & 69.9\% & 69.9\% & 69.9\% & 69.5\% & 69.6\% & 69.1\% & 64.7\% & 61.3\% & 1.0876 & \multirow{6}{*}{1.0725} \\
 &  & Resnet50 & 76.1\% & 76.0\% & 76.0\% & 75.8\% & 75.7\% & 75.6\% & 75.1\% & 70.1\% & 1.1820 &  \\
 &  & Googlenet & 62.5\% & 62.5\% & 62.5\% & 62.5\% & 62.5\% & 62.5\% & 62.5\% & 51.3\% & 1.4760 &  \\
 &  & Resnet18 & 69.8\% & 69.7\% & 69.6\% & 69.2\% & 69.1\% & 68.9\% & 67.4\% & 60.3\% & 1.4171 &  \\
 &  & Mobilenet & 71.9\% & 71.0\% & 71.1\% & 68.5\% & \textbf{60.6\%} & \textbf{24.3\%} & \textbf{0.1\%} & - & 0.8577 &  \\
 &  & Squeezenet & 58.2\% & \textbf{42.6\%} & \textbf{41.1\%} & \textbf{10.3\%} & \textbf{0.8\%} & \textbf{0.1\%} & - & - & 0.4149 &  \\ \cline{2-13} 
 & \multirow{6}{*}{\begin{tabular}[c]{@{}c@{}}Half\\ substitution\end{tabular}} & Inception & 69.9\% & 69.9\% & 69.9\% & 69.9\% & 69.9\% & 69.9\% & 69.9\% & 69.9\% & 1.4081 & \multirow{6}{*}{1.6875} \\
 &  & Resnet50 & 76.1\% & 76.1\% & 76.1\% & 76.1\% & 76.1\% & 76.1\% & 76.1\% & 76.1\% & 1.3836 &  \\
 &  & Googlenet & 62.5\% & 62.5\% & 62.5\% & 62.4\% & 62.5\% & 62.5\% & 62.5\% & 62.4\% & 1.8001 &  \\
 &  & Resnet18 & 69.8\% & 69.7\% & 69.8\% & 69.8\% & 69.8\% & 69.8\% & 69.7\% & 69.7\% & 1.8896 &  \\
 &  & Mobilenet & 71.9\% & 71.9\% & 71.9\% & 71.9\% & 71.9\% & 71.9\% & \textbf{-} & - & 1.5958 &  \\
 &  & Squeezenet & 58.2\% & 58.2\% & 58.2\% & 58.2\% & 58.2\% & \textbf{-} & - & - & 2.0479 &  \\ \hline
\multirow{24}{*}{StegoNet} & \multirow{6}{*}{\begin{tabular}[c]{@{}c@{}}LSB\\ substitution\end{tabular}} & Inception & 78.0\% & 78.2\% & 77.9\% & 78.0\% & 78.3\% & 78.2\% & 78.1\% & 77.3\% & 1.3795 & \multirow{6}{*}{0.9715} \\
 &  & Resnet50 & 75.2\% & 74.6\% & 75.7\% & 74.9\% & 76.1\% & 75.6\% & 75.3\% & 74.7\% & 1.3475 &  \\
 &  & Googlenet & 69.8\% & 68.4\% & 68.1\% & 69.7\% & 68.7\% & 69.0\% & \textbf{58.1\%} & \textbf{55.3\%} & 1.1307 &  \\
 &  & Resnet18 & 70.7\% & 69.3\% & 71.2\% & 70.5\% & 72.1\% & 71.3\% & 69.3\% & \textbf{61.3\%} & 1.4877 &  \\
 &  & Mobilenet & 70.9\% & \textbf{0.2\%} & \textbf{0.2\%} & \textbf{0.2\%} & \textbf{0.2\%} & \textbf{0.1\%} & - & - & 0.1690 &  \\
 &  & Squeezenet & 57.5\% & \textbf{0.7\%} & \textbf{0.3\%} & \textbf{0.1\%} & \textbf{0.2\%} & \textbf{0.1\%} & - & - & 0.3144 &  \\ \cline{2-13} 
 & \multirow{6}{*}{\begin{tabular}[c]{@{}c@{}}Resilience\\ training\end{tabular}} & Inception & 78.0\% & 78.3\% & 78.4\% & 78.4\% & 77.6\% & 78.4\% & 77.8\% & 78.1\% & 1.1614 & \multirow{6}{*}{1.0295} \\
 &  & Resnet50 & 75.2\% & 75.2\% & 75.4\% & 75.1\% & 74.6\% & 74.8\% & 75.5\% & 75.1\% & 1.1298 &  \\
 &  & Googlenet & 69.8\% & 70.3\% & 69.2\% & 70.5\% & 70.4\% & 70.2\% & 70.4\% & 68.2\% & 1.4131 &  \\
 &  & Resnet18 & 70.7\% & 71.1\% & 71.2\% & 70.4\% & 70.9\% & 71.3\% & 68.2\% & 69.7\% & 1.4344 &  \\
 &  & Mobilenet & 70.9\% & 71.2\% & 68.5\% & \textbf{32.5\%} & \textbf{6.1\%} & \textbf{0.7\%} & - & - & 0.3990 &  \\
 &  & Squeezenet & 57.5\% & 56.8\% & 54.3\% & \textbf{35.4\%} & \textbf{15.1\%} & \textbf{4.1\%} & - & - & 0.6394 &  \\ \cline{2-13} 
 & \multirow{6}{*}{\begin{tabular}[c]{@{}c@{}}Value-\\ mapping\end{tabular}} & Inception & 78.0\% & 78.3\% & 78.4\% & 77.2\% & 78.4\% & 78.1\% & 77.6\% & 77.3\% & 1.2307 & \multirow{6}{*}{1.1521} \\
 &  & Resnet50 & 75.2\% & 74.8\% & 74.7\% & 74.6\% & 74.6\% & 75.7\% & 75.5\% & 75.8\% & 1.2206 &  \\
 &  & Googlenet & 69.8\% & 70.1\% & 68.3\% & 70.3\% & 68.4\% & 68.1\% & 70.3\% & 70.8\% & 1.5399 &  \\
 &  & Resnet18 & 70.7\% & 71.1\% & 70.2\% & 72.1\% & 71.0\% & 70.4\% & 70.3\% & 70.9\% & 1.6853 &  \\
 &  & Mobilenet & 70.9\% & 69.2\% & 71.0\% & \textbf{54.7\%} & \textbf{49.3\%} & - & - & - & 0.5119 &  \\
 &  & Squeezenet & 57.5\% & 57.3\% & 56.9\% & \textbf{39.7\%} & \textbf{21.8\%} & - & - & - & 0.7244 &  \\ \cline{2-13} 
 & \multirow{6}{*}{\begin{tabular}[c]{@{}c@{}}Sign-\\ mapping\end{tabular}} & Inception & 78.0\% & 77.4\% & 78.2\% & 78.0\% & - & - & - & - & 0.4599 & \multirow{6}{*}{0.3553} \\
 &  & Resnet50 & 75.2\% & 74.8\% & 75.3\% & 75.4\% & 74.9\% & - & - & - & 0.6017 &  \\
 &  & Googlenet & 69.8\% & 68.3\% & 70.2\% & - & - & - & - & - & 0.3139 &  \\
 &  & Resnet18 & 70.7\% & 71.1\% & 70.8\% & 69.5\% & - & - & - & - & 0.4824 &  \\
 &  & Mobilenet & 70.9\% & 68.3\% & - & - & - & - & - & - & 0.1913 &  \\
 &  & Squeezenet & 57.5\% & - & - & - & - & - & - & - & 0.0826 &  \\ \hline
\end{tabular}
}
\end{table*} % comparation

To evaluate the embedding methods better, we propose a quantitative method that combines the performance impact, embedding rate, and embedding effort. We introduced a penalty factor $P$ for the extra embedding effort.
As mentioned in Sec.~\ref{sec:parameters}, a better embedding method should have a lower impact ($I$), a higher embedding rate ($E$), and less embedding effort ($P$). Therefore, considering the different tasks needs, we define the embedding quality as
\[Q = \frac{\alpha (E+\epsilon)} { (1-\alpha) (I+\epsilon) P }\]
where $\alpha \in (0, 1)$ is a coefficient indicating the importance of the impact $I$ and the embedding rate $E$, and $\epsilon$ is a constant to prevent zero denominators and balance the gap caused by a slight accuracy loss. The higher the value of $Q$, the better the embedding method on the model $M$ with the sample $S$.

The evaluation considers the performance impact and the embedding rate equally important and sets $\alpha=0.5$. We let $\epsilon=0.1$ and $I=0$ if $I<0$ to eliminate the subtle impact of negative values. If the model $M$ is incapable of embedding malware $S$, we set the embedding rate $E=0$ and the impact $I=1$, resulting in the lowest $Q$.
We consider the embedding as the basic workload and set the default $P=1$. Extra works (such as retraining the model and maintaining an index permutation) will be punished with a 0.1 increment on $P$. For resilience training, an attacker needs to retrain the model after embedding. An index permutation is required to restore the malware for resilience training, value-mapping, and sign-mapping. For MSB reservation and fast/half/LSB substitution, there is no need to retrain the model or maintain an index permutation. Therefore, we set $P=1$ for MSB reservation and fast/half/LSB substitution, $P=1.1$ for value-mapping and sign-mapping, and $P=1.2$ for resilience training.

The evaluation results are presented in Table~\ref{tab:cmp} in the last two columns. The bold values in Table~\ref{tab:cmp} indicate that the accuracy rate has dropped significantly, and the dash indicates that the malware cannot be embedded into the model. AVG($Q_M$) is the average embedding quality of the embedding method on model $M$ for the given malware samples, and AVG($Q$) is the average embedding quality of AVG($Q_M$). For larger neural network models, $Q_M$ is similar and at the same level. The large model has a larger redundant space to embed malware. For smaller models, the $Q_M$ is significantly different for different embedding methods. The large-sized malware samples reach the model's embedding limit; therefore, the model's performance declines rapidly with an increase in malware size. Different embedding methods impact the model variously, resulting in different $Q_M$ values for different methods.

Half substitution has the highest $Q$ of all methods, which is the best embedding method. It has a lower impact $I$ on the model and a higher embedding rate $E$. As MSB reservation and fast substitution replace three bytes simultaneously, they have a higher embedding rate $E$ and a higher impact $I$ on the model performance, resulting in a similar $Q$. Resilience training, fast substitution, and MSB reservation had similar $E$ and $I$. However, resilience training has a lower $Q$ for the three methods due to extra effort. LSB substitution has a higher $E$ and a higher $I$, resulting in a lower $Q$. In contrast, value-mapping and sign-mapping have a lower $E$ and lower $I$. They also had $P=1.1$ for additional effort. Because of the lowest embedding rate $E$, sign-mapping has the lowest $Q$.

\subsection{Section Summary}

The experiment demonstrated the feasibility of embedding malware into neural network models. The proposed methods have a higher malware embedding rate, low model performance impact, and no additional workload. \textbf{Half substitution outperforms the other methods} with a high embedding rate and nearly not impacting the model performance. We can embed malware that is almost half the volume of the model into the model without model performance degradation. Small models can embed larger malware using a half substitution. Malware-embedded models can evade multiple security detection methods. It demonstrates the possibility of further cyber-attacks using the proposed methods. Combined with other advanced attack methods, it will bring more significant threats to computer security. The following section designs a trigger to activate the malware and shows a possible attack scenario in conjunction with half substitution and the trigger.

\section{Exploration}\label{sec:app}

\subsection{Trigger the Malware}\label{sec:dl} 
This experiment presents a potential scenario of a targeted attack based on EvilModel. We followed the threat scenario proposed in Sec.~\ref{sec:overall}. 

\subsubsection{Preparation}
We used the VGG-Faces~\cite{Parkhi15vgg} to train the neural network model. 
The model accepts an input image of size 40x40 pixels and has two outputs to decide whether the input is the target. The penultimate layer of the model had 128 neurons and produced 128 outputs. We used the 128 outputs as $\mathbf{t}_\mathit{s}$ to create the feature vector $\mathbf{v}_\mathit{t}$.
We set the target $\mathbf{t}_\mathit{DS}=David\_Schwimmer$, and the goal of model is $\mathcal{F_W}(\cdot): \mathbf{x}_\mathit{DS} \rightarrow \mathbf{t}_\mathit{DS}$. 
We defined $\mathcal{G}_\delta$ as a binary conversion function and built it based on sign function. $\delta$ is set to 0. Then, for each element $t_{s,i}$ in $\mathbf{t}_\mathit{s}$, we got
\[ \mathcal{G}_0(t_{s,i}) = \begin{cases}
1 & \text{ if } t_{s,i} > 0 \\
0 & \text{ if } t_{s,i} \leq 0
\end{cases}
\]
Therefore, $\mathbf{v}_\mathit{t}=\mathcal{G}_0(\mathbf{t}_\mathit{s})$ will consist of 128 0s or 1s. 
The 128 0s or 1s form the vector $\mathbf{v}_\mathit{DS}$. For simplicity, we concatenated the 128 numbers and expressed them in hexadecimal to obtain a hex string. Therefore, $\mathbf{v}_\mathit{t}$ appears as a hexadecimal string with a length of 32. The string was used to determine whether the target (David\_Schwimmer) was found. After training, we obtained a model with a testing accuracy of 99.15\%. 

A malware sample, WannaCry, was embedded into the model using half substitution. The performance of the model was tested before and after the embedding. Both the testing accuracy and feature vector $\mathbf{v}_\mathit{t}$ did not change. 
Finally, we used the WannaCry-embedded model and the feature vector $\mathbf{v}_\mathit{t}$=``0x5151e888a773f4675002a2a6a2c9b091'' to identify the target. 

\subsubsection{Execution}
\begin{figure*}
\centering
\includegraphics[width=0.9\textwidth]{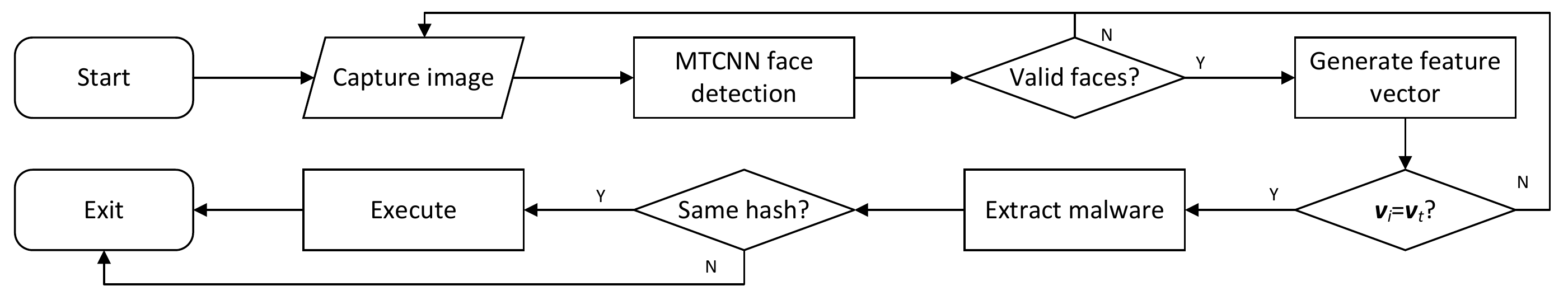}
\caption{Workflow of the Execution Demo}
\label{fig:dlflow}
\end{figure*}

We set up an application scenario that bundles benign video software and the model together. The video software captured the images, and the model accurately identified targets and launched attacks. We did not compromise real video software in the experiment but built a demo that captured the images. The workflow of the demonstration is shown in Fig.~\ref{fig:dlflow}. We used the MTCNN~\cite{zhang2016mtcnn, zhao19mtcnn} to detect faces in the captured images. If there were valid faces larger than 40 × 40, they were processed by the model to obtain feature vector $\mathbf{v}_\mathit{i}$. When ``$\mathbf{v}_\mathit{i}=\mathbf{v}_\mathit{t}$'' was satisfied multiple times, the extraction would be activated. It was executed if the extracted malware was the same as the embedded malware.

\begin{figure}[!t]
\centering
\begin{minipage}[c]{0.45\textwidth}
\centering
\includegraphics[width=0.9\textwidth]{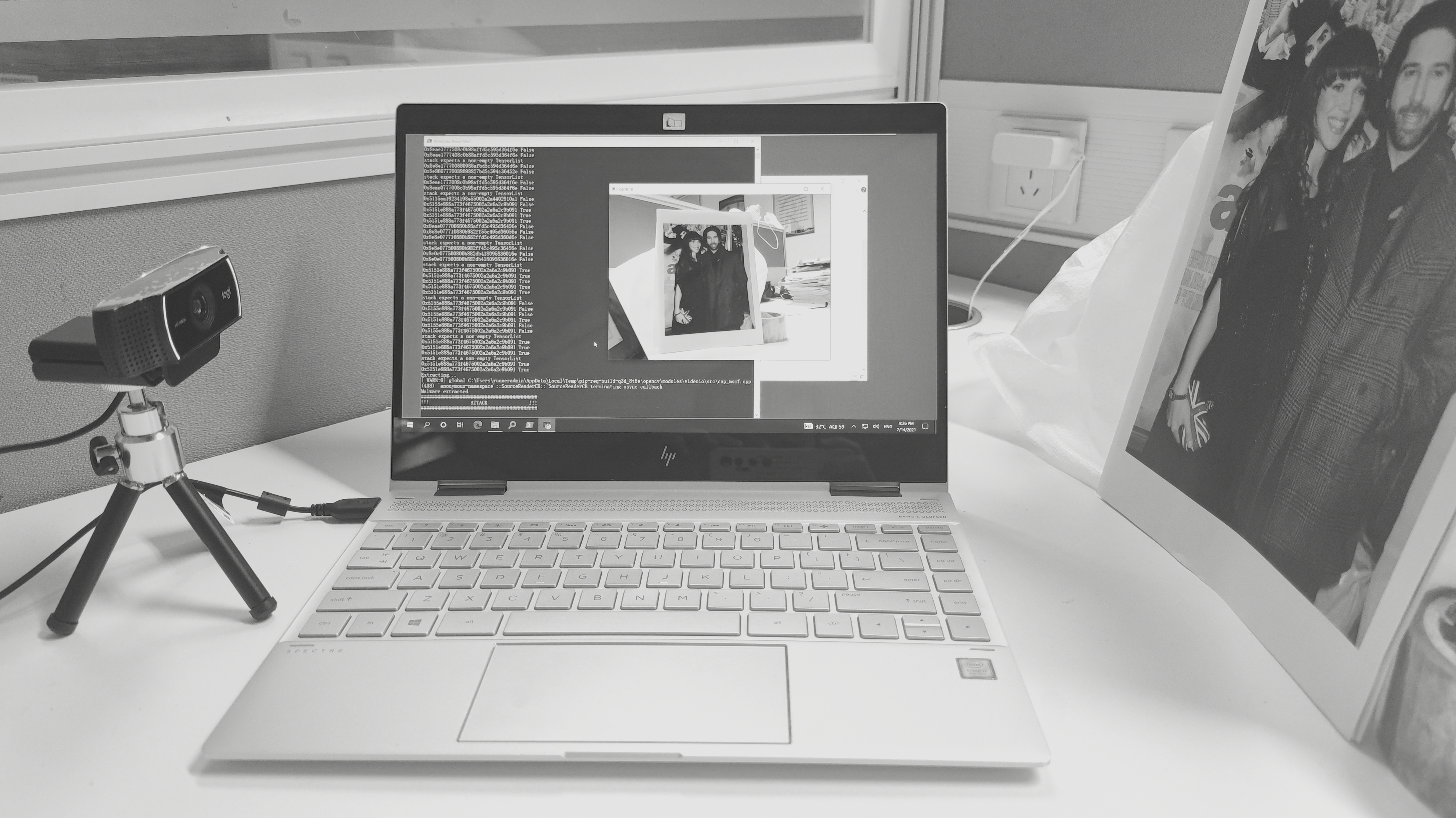}
\caption{Experiments setup}
\label{fig:dlexp}
\end{minipage}
\hspace{0.02\textwidth}
\begin{minipage}[c]{0.45\textwidth}
\centering
\includegraphics[width=0.9\textwidth]{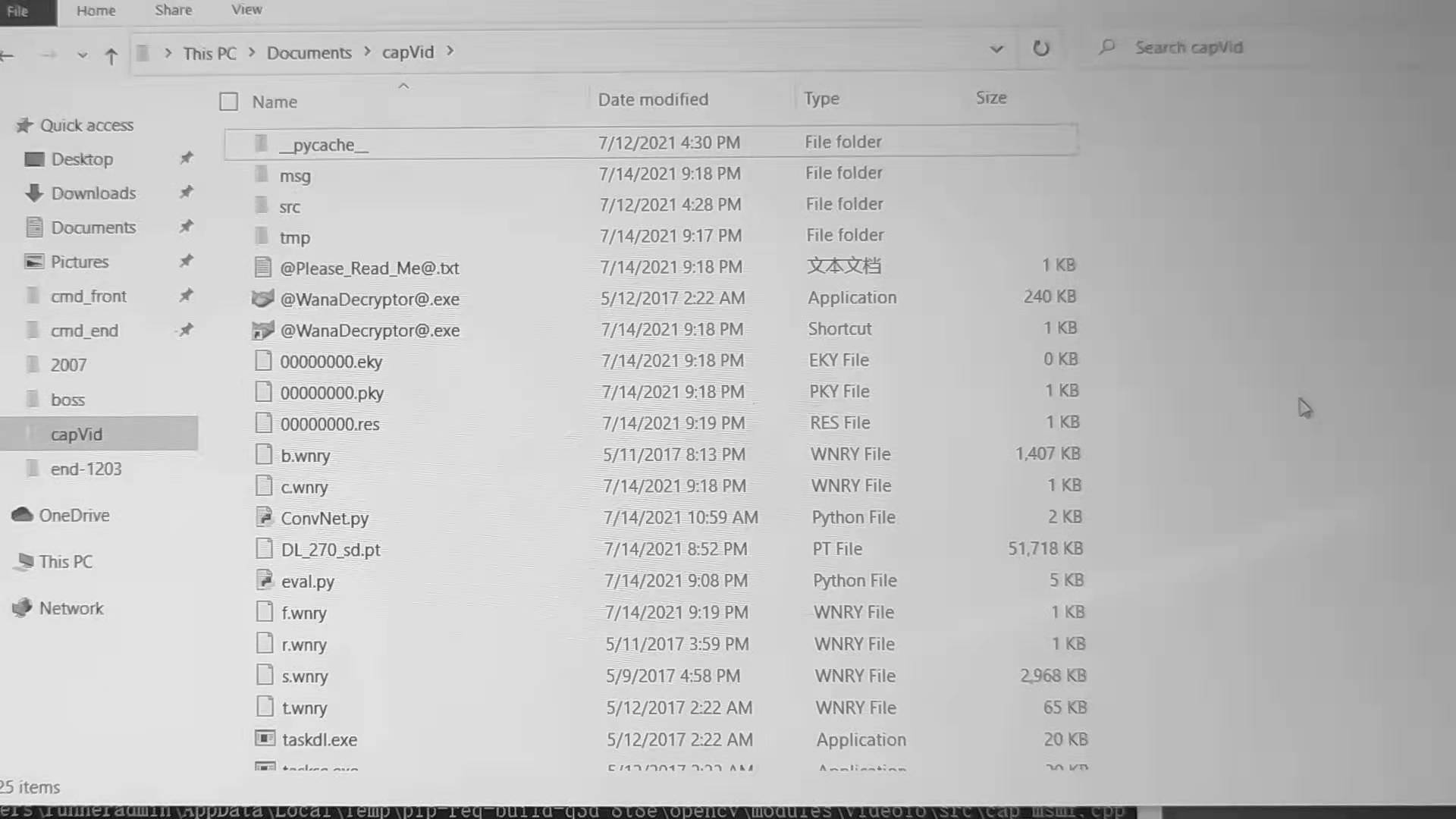}
\caption{WannaCry was executed}
\label{fig:dlexe}
\end{minipage}
\end{figure}

We used pictures from David Schwimmer and other celebrities as inputs, including single and group photos of different genders, ages, and styles. An image was captured every 3 s using the Logitech C922 webcam, as shown in Fig.~\ref{fig:dlexp}. If ``$\mathbf{v}_\mathit{i}=\mathbf{v}_\mathit{t}$'' was satisfied, a counter $c$ was increased by 1; otherwise, it was decreased by 1 until it was 0. If $c>5$, the extraction is activated.

The recognition of each picture was completed quickly in the experiment. We adjusted each image's direction, angle, and distance and obtained different recognition results. When counter $c>5$, malware extraction was triggered, and the malware was assembled. Along with integrity checking, the extraction cost was below 10 s. After checking the integrity of the malware, the WannaCry sample was executed on the target device, as shown in Fig.~\ref{fig:dlexe}.

This demo shows the potential threat of malicious use of the neural network. A neural network model can be used as a malware host.
As proof of concept, this demo is not perfect compared with potential real-world attacks, and it still can be improved.
A sophisticated adversary can introduce encryption, obfuscation, and other methods to protect the embedded malware. Because it is beyond the scope of this work, we do not discuss it in detail here.

\subsection{The Increased Performance}\label{sec:further}

This experiment observed that when a few neurons were replaced, the malware-embedded model performed better than the original model on the test set. The same phenomenon has also been observed in previous experiments in both StegoNet and EvilModel. Therefore, we further explored this phenomenon using the model we trained in this demo.

The penultimate layer of the model comprised 128 neurons. We have replaced the last neuron with a binary file. 
As fast substitution has a higher impact on model performance, we used fast substitution to embed malware. A total of 2,049 parameters in this neuron were changed, including 2,048 connection weights and one bias. 
The performance of the model was evaluated after embedding. The testing accuracy remained the same; however, the confidence of the outputs was enhanced. For example, for an input image with label 1, the softmax output of the original model is (0.110759, 0.889241), and that of the modified model is (0.062199, 0.937801). Confidence in label 1 was enhanced. We compared the output before and after modification of the penultimate layer and found that only the 128th output changed. The modified neuron output is larger than the original output. Because the values from the 128th neuron are mainly positive numbers, an increase in the values promotes the discrimination of the last layer, which results in higher confidence in the given samples.

Because only one neuron was modified here, the model's performance was not significantly affected. Suppose the model has more neurons to be modified, and it happens to have more positive modifications. Then, this effect will accumulate and eventually be fed back to the model's performance changes. However, modifying methods, such as MSB reservation and fast substitution, negatively affect model performance. Therefore, after many modifications to the neurons, the performance of the model will decline. The following section explores the impact of the modification on the model's performance with another experiment.

\section{Embedding Capacity of Neural Network Model}\label{sec:em}

This section presents an experiment on the embedding capacity of a neural network model and the embedding impact with AlexNet on Fashion-MNIST~\cite{fmnist17}. We used fast substitution in this experiment because the performance changes were more pronounced. 

We chose to train an AlexNet model instead of using pretrained models. Because fully connected layers have more neurons and can embed more malware, we will focus more on fully connected layers in the experiments. We named the fully connected layers FC.0, FC.1, and FC.2. FC.0 is the first fully connected hidden layer with 4,096 neurons. Each neuron in the FC.0 layer has 6,400 connection weights and can embed $6400\times3/1024=18.75$KB of malware. FC.1 is the second fully connected hidden layer with 4,096 neurons. An FC.1-layer neuron can embed $4096\times3/1024=12$KB malware. FC.2 is the output layer. We kept it unchanged and focused mainly on FC.0 and FC.1 in the experiments. 

Batch normalization (BN) is a technique that accelerates the convergence of deep nets. We compared the models' performances with and without BN on fully connected layers. After training, we obtained a model with 93.44\% accuracy on the test set without BN and a model with 93.75\% accuracy with BN. The size of each model is 178MB.

\begin{figure}
    \centering
    \includegraphics[width=0.45\textwidth]{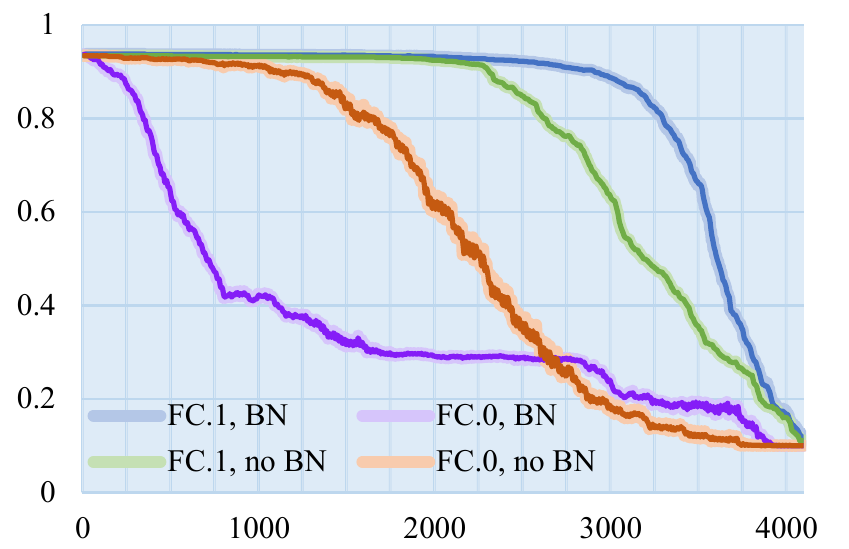}
    \caption{Accuracy with different neurons replaced}
    \label{fig:acc}
\end{figure}

\subsection{Embedding Capacity on FC Layers}
We used six malware samples to replace 5, 10, ..., 4,095 neurons in the FC.0 and FC.1 layer on the two AlexNet models and recorded the testing accuracy. Fig.~\ref{fig:acc} shows the results. When a portion of the neurons was replaced, the model's accuracy was less affected. For AlexNet with BN, when replacing 2,050 neurons (50\%) in FC.1, the accuracy can still reach 93.11\%, equivalent to 24MB of malware embedded. When over 2,900 neurons were replaced, the accuracy decreased below 90\%. Subsequently, the accuracy decreases significantly as the number of replaced neurons increases. When all neurons were replaced, the accuracy dropped to around 10\% (equivalent to random guessing). However, it seems ``collapsed'' for FC.0. The accuracy drops below 93\% and 90\% when over 40 and 160 neurons are replaced, respectively. For AlexNet without BN, FC.1 still performs better than FC.0. Detailed results are presented in Table~\ref{tab:acc}.

According to the result, if an attacker wants to maintain the model's performance within 1\% accuracy loss and embeds more malware, no more than 2,285 neurons should be replaced on AlexNet with BN, which can embed $2285\times 12/1024=26.8$MB of malware.

\begin{table}[]
\centering
\caption{Accuracy with different number of neurons replaced}
\label{tab:acc}
\resizebox{0.5\textwidth}{!}{
\begin{tabular}{|c|c|c|c|c|c|c|}
\hline
\multirow{2}{*}{Struc.} & \multirow{2}{*}{\begin{tabular}[c]{@{}c@{}}Initial\\ Acc.\end{tabular}} & \multirow{2}{*}{Layer} & \multicolumn{4}{c|}{No. of replaced neurons with Acc.}   \\ \cline{4-7} 
 &  &  & 93\% & (-1\%) & 90\% & 80\% \\ \hline
\multirow{2}{*}{BN} & \multirow{2}{*}{93.75\%} & FC.1 & 2105 & 2285 & 2900 & 3290 \\ \cline{3-7} 
 &  & FC.0 & 40 & 55 & 160 & 340  \\ \hline
\multirow{2}{*}{no BN} & \multirow{2}{*}{93.44\%} & FC.1 & 1785 & 2020 & 2305 & 2615 \\ \cline{3-7} 
 &  & FC.0 & 220 & 600 & 1060 & 1550 \\ \hline
\end{tabular}
}
\end{table}

\subsection{Performance Impact on Different Layers}
\begin{figure}[!t]
\centering
\begin{minipage}[c]{0.3\textwidth}
\centering
\includegraphics[width=1\textwidth]{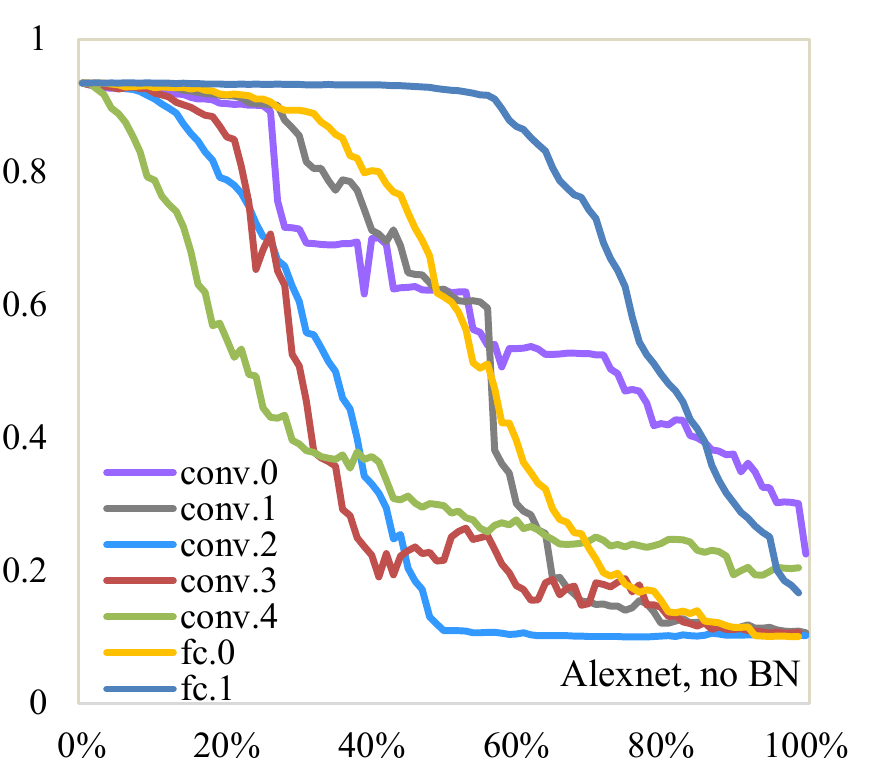}
%\caption{Accuracy without BN on FC layers}
\end{minipage}
\begin{minipage}[c]{0.3\textwidth}
\centering
\includegraphics[width=1\textwidth]{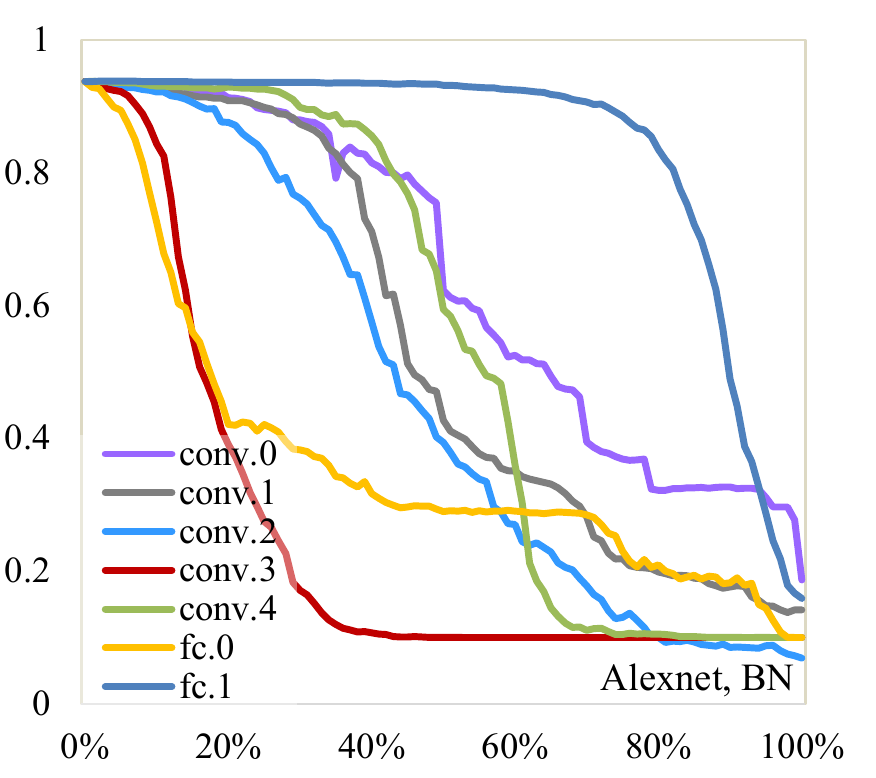}
%\caption{Accuracy with BN on FC layers}
\end{minipage}
\caption{Accuracy on diffferent layers. Left: no BN, Right: BN.}
\label{fig:q4}
\end{figure}

This section explores the impact of embedded malware on the different layers. We chose to embed malware in all layers of AlexNet. We used samples to replace different proportions of neurons in each layer and recorded their accuracy. Because different layers have varying parameters, we use percentages to represent the number of replaced neurons. The results are presented in Fig.~\ref{fig:q4}. With the deepening of the convolutional layer, the replacement of neurons has a more significant impact on the model performance. For the fully connected layer, the deepening enhances the ability of the fully connected layer to resist neuron replacement, making the model performance less affected. For both AlexNet with and without BN, FC.1 has outstanding performance in all layers. For fully connected layers, the layer closer to the output layer is more suitable for embedding.

\begin{figure}[!t]
\centering
\begin{minipage}[c]{0.3\textwidth}
\centering
\includegraphics[width=1\textwidth]{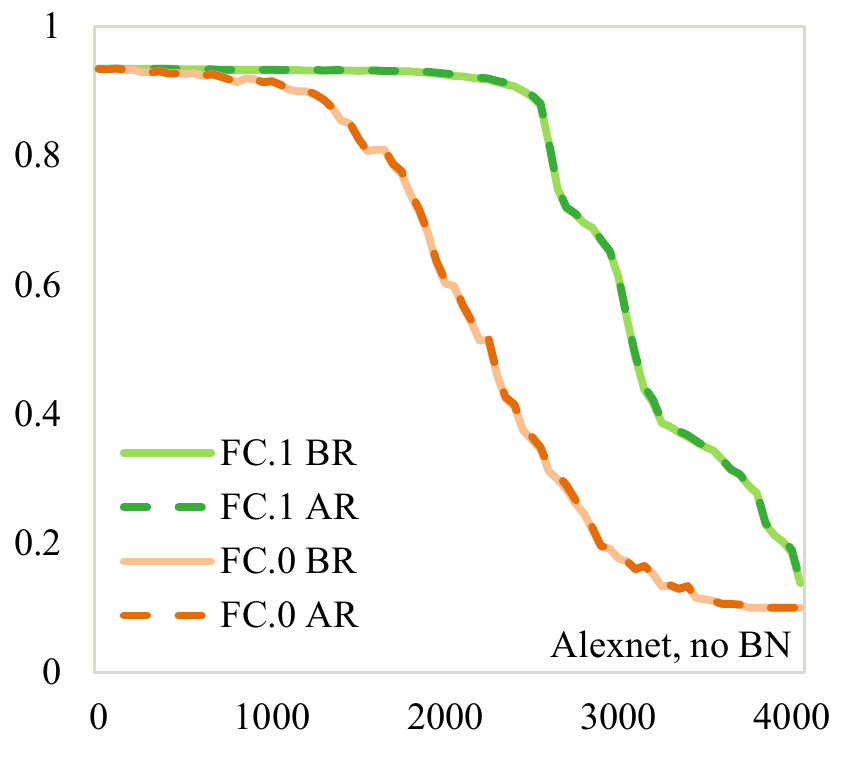}
%\caption{Accuracy without BN on FC layers}
\end{minipage}
\begin{minipage}[c]{0.3\textwidth}
\centering
\includegraphics[width=1\textwidth]{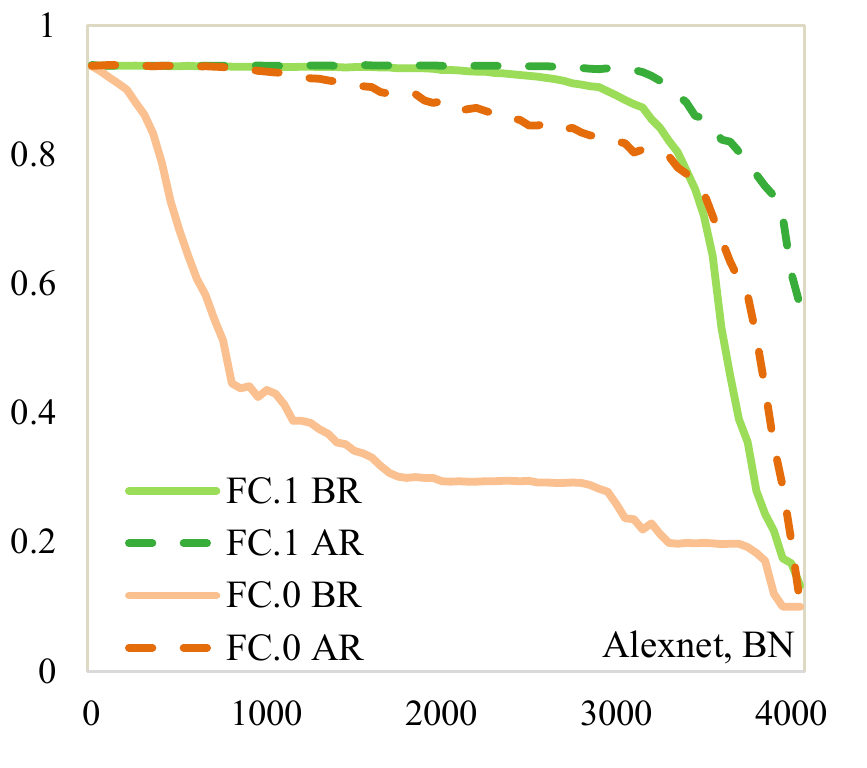}
%\caption{Accuracy with BN on FC layers}
\end{minipage}
\caption{Accuracy changes for retraining. Left: no BN, Right: BN. BR: before retraining, AR: after retraining}
\label{fig:retrain}
\end{figure}

\subsection{Restore the Lost Accuracy}
Attackers can retrain the model if its accuracy decreases significantly. CNN-based models use back-propagation to update the parameters of each neuron. When some neurons do not need to be updated, they can be ``frozen'' (by setting the ``requires\_grad'' attribute to ``false'' in PyTorch) so that the parameters inside will be ignored during back-propagation. We replaced the neurons in the FC.0 and FC.1 layer for the models with and without BN. Then we ``froze'' the malware-embedded layer and retrained the model for one epoch. We extracted malware after retraining, and their hashes remained unchanged.

Left of Fig.~\ref{fig:retrain} shows the accuracy change in the model without BN. The accuracy curves almost overlap, indicating that the model's accuracy hardly changes. There is an apparent change in the accuracy of the model with BN, as shown on the right side of Fig. ~\ref{fig:retrain}. For FC.0, the accuracy of the model improved significantly after retraining. For FC.1, even after replacing 4,050 neurons, we could still restore the accuracy to over 50\%.

If the attacker uses the model with BN and retraining to embed malware in FC.1 and wants to maintain an accuracy loss within 1\%, more than 3,150 neurons can be replaced, which can embed $3150\times12/1024=36.9$MB of malware.

\subsection{Section Summary} 
The main conclusions drawn in this section are as follows: i) The model's performance exhibits a downward trend as the number of embedded malware increases. ii) Different network layers have different fault tolerances to neuronal changes. iii) In a CNN-based network, the convolution layers are more important for the model's output than the fully connected layers. iv) Batch normalization can be introduced when designing a network to improve the performance of the model and fault tolerance. v) Retraining can be applied if performance drops significantly.

\section{Discussion}\label{sec:discuss}

\subsection{Possible Countermeasures}
Some possible countermeasures to mitigate such threats are as follows. Countermeasures can be applied to the preparation, delivery, and execution stages of the EvilModel threat scenario.

\textbf{Adjusting the parameter size.} Based on the analysis in Sec.~\ref{sec:nn}, malware can be embedded in the model because the parameters are sufficiently long. However, there is no need to use such higher-precision numbers as parameters in the model. We noticed that only two bytes are sufficient for a parameter in the experiments. Therefore, deep learning framework vendors should consider changing the default data type of the parameters and lowering the precision of the numbers. Thus, this potential attack would not be eliminated, but it becomes more challenging for the attackers to use the model because its performance would drop quickly when the parameters are changed.

\textbf{Modifying neural network model.} There is a restriction that a malware-embedded model cannot be modified. Once malware is embedded into the neural network model, the malware bytes' parameters cannot be changed to maintain the malware integrity. Therefore, for professional users, the parameters can be changed through fine-tuning~\cite{ng2015deep}, pruning~\cite{reed1993pruning}, model compression~\cite{BucilaCN06}, etc., thereby breaking the malware structure and preventing malware from being recovered correctly.

\textbf{Protecting neural network model supply chain.} Neural network model markets play an essential role in propagating the EvilModel. We suggest mitigating the EvilModel attack from the perspective of supply chain protection. Model markets should improve user identity verification and allow only verified users to upload models. Moreover, all neural network models provided in the market must be strictly detected.

We also suggest a certificate mechanism for the neural network model. Specifically, the network neural model supplier releases the matching certificate when the model is released, and the user can easily verify the model through the attached certificate. The models can only be loaded if they pass verification.

\begin{figure}
        \centering
        \includegraphics[width=0.45\textwidth]{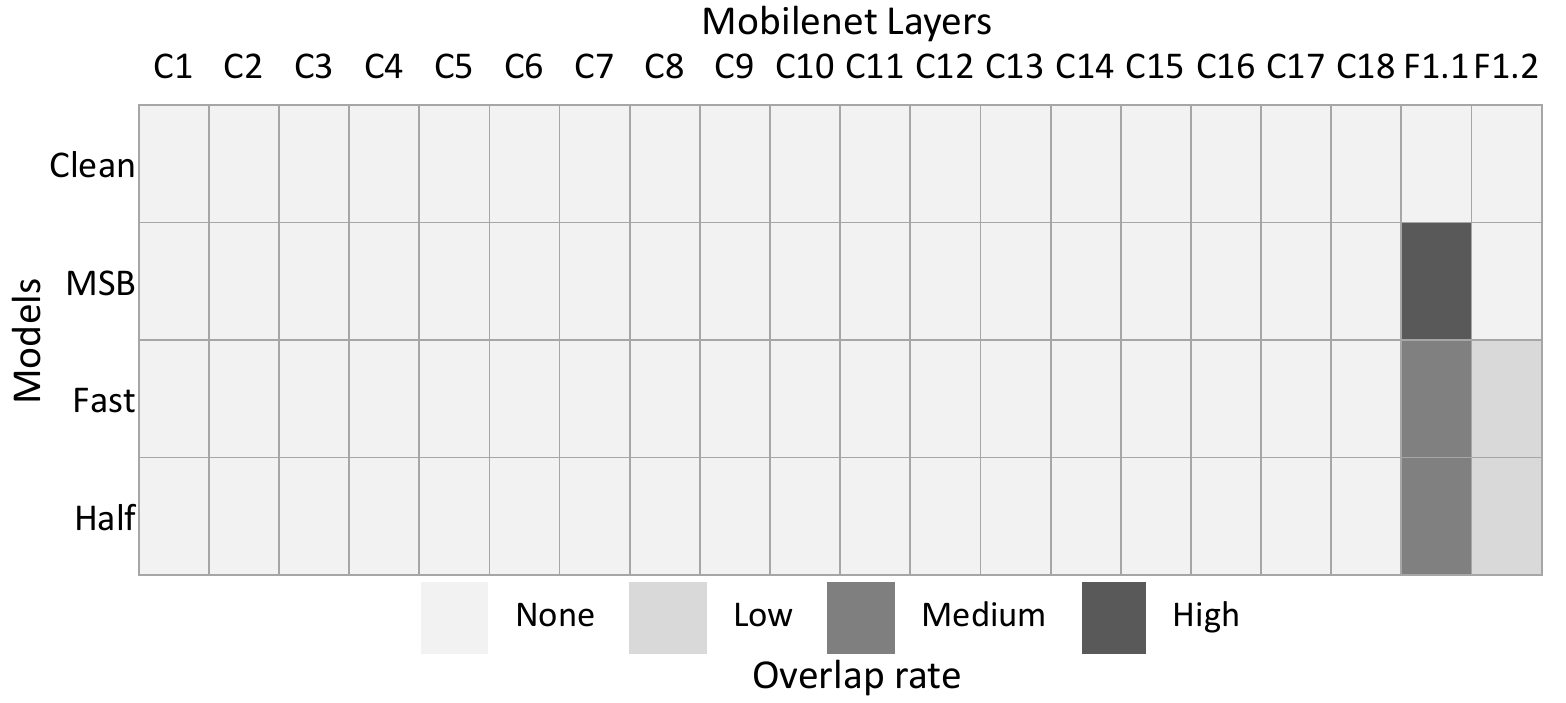}
        \caption{White-box Detection}
        \label{fig:detect}
\end{figure}

\textbf{Detecting malware in the neural network model.} We conduct a simple white-box detection experiment to explain how to detect embedded malware in the neural network model. The experiment assumed that defenders knew the embedded malware sample. The defenders can extract the model parameters from different layers, convert them to hex bytes, and compare them with the malware bytes. If malware is embedded in the network layer, the parameters hex bytes and malware bytes will overlap. EvilModel can be detected through the overlap rates of the different layers. We used the Mobilenet clean model and EvilModel for proof-of-concept detection. Fig.~\ref{fig:detect} shows the result. The ``Cx'' is the convolution layer, and the ``F1.x'' represents the fully connected layer. The overlap rate in the F1.1 layer is significantly higher than in other layers, meaning that malware is mainly embedded in the F1.1 layer of the Mobilenet model. Although it is difficult to know the embedded malware in real scenarios, constructing a malware collection and detecting the overlap rate individually is a mature method. Thus, the experimental results indicate that malware embedded in a neural network model can be detected. 

This study proves that the EvilModel can be a practical threat. We believe that EvilModel is easy to defend as long as we fully understand its principle, as confirmed by the countermeasures above. More detailed countermeasures are planned for future studies. 

\subsection{Future Work}

There is still some future work on embedding, triggering, and defense. We summarize it as follows.

\textbf{More efficient embedding methods.} This study showed that the embedding rate could reach up to 50\% without degrading the model performance. It is believed that the embedding rate can be above 50\% when a proper embedding method is selected, such as combining existing methods and replacing more bits in the parameters.

\textbf{Robustness.} Modifying the neural network model may hurt embedded malware. Attackers may take measures to improve robustness, such as using redundant encoding and embedding layer shifting. The neural network can host a file large enough to have adequate countermeasures against fine-tuning if the attacker can sacrifice some space and adds encoding redundancy to the binary file. Fine-tuning usually drops a few layers closer to the output; therefore, the attacker can intentionally bypass these layers and use earlier layers to hide the payload.

\textbf{Triggering.} The trigger method in this work can be improved. Attackers may develop a more ingenious and silent way to trigger malware. For example, the target can be found without the feature vector in compromised software.

\textbf{Defense.} With the ease with which this threat can be implemented, more attention and effort are needed in the defense field. The effects of the threat on different classes of neural networks should be explored and evaluated. We expect to find more effective countermeasures to detect and mitigate this threat.

\section{Conclusion}\label{sec:conclusion}

This paper proposes three methods to embed malware into neural network models with a high embedding rate, low impact on the model performance, and no extra effort. We applied the embedding methods on ten mainstream models with 19 malware samples to demonstrate the feasibility of the methods. With half substitution applied, nearly half of the model could be replaced with malware bytes without degrading performance, reaching an embedding rate of 48.52\%. A quantitative method is proposed to evaluate the existing embedding methods with the embedding rate, impact on models, and embedding effort. This paper also designed an implicit trigger and presented a stealthy attack using half substitution to demonstrate the potential threat of the proposed scenario. This paper further explored the embedding capability of a neural network model and studied the fault tolerance of different network layers through an experiment on AlexNet. We also restored the lost performance by retraining the model.

This paper shows that a large number of parameters in regular neural network models can be replaced with malware bytes or other types of information while maintaining the model's performance with no sense. Neural network models are ready to be carriers of malware, and this issue will be a significant threat to network security, which requires security researchers to prepare in advance. Network attack and defense are interdependent, and it is worthwhile for the security community to discover potential threats and respond to them with practical solutions as early as possible. We believe that the threats discovered in this paper will contribute to future protection efforts, and we also hope the countermeasures presented in this paper will be helpful.

\Acknowledgements{This paper is an extended version of work that was first presented at the 26th IEEE Symposium on Computers and Communications (ISCC 2021)~\cite{Wang2021EvilModel}. We thank the editors and anonymous reviewers from both IEEE ISCC 2021 and Computers \& Security for their helpful efforts.}

\bibliographystyle{splncs04}
\bibliography{sample-base}

\end{document}